\documentclass[aps,twocolumn,nofootinbib]{revtex4}
\usepackage{adjustbox}
\usepackage{graphicx} 
\usepackage{caption,subcaption}
\usepackage{amsmath}
\usepackage{amsthm}
\usepackage{framed}
\usepackage{tikz}

\usepackage{comment}
\usepackage{bbm}
\usepackage[version = 4]{mhchem}
\usepackage{tikz-cd} 
\usepackage{tikz}
\usepackage{amssymb}
\usepackage{mathtools}

\usepackage{amsthm}

\theoremstyle{plain}%default 
\newtheorem{theorem}{Theorem}[section]

\theoremstyle{definition} 

\newtheorem{example}{Example}[section]

\theoremstyle{remark} 
\usepackage{physics}
\usepackage{xcolor}
\usepackage{circledsteps}
\usepackage{bm}
\usepackage{mathtools}
\usepackage[hidelinks]{hyperref}
\usepackage{algorithm}
\usepackage{algpseudocode}
\usepackage{xcolor}
\usepackage{xfrac}
\usepackage{bm}
\usepackage[T1]{fontenc}

\usepackage{xfrac} % nice fractions

\usepackage{upquote}

\DeclareMathOperator{\Spc}{\mathcal{S}}
\DeclareMathOperator{\C}{\mathcal{C}}
\DeclareMathOperator{\Cpx}{\mathfrak{C}}
\DeclareMathOperator{\R}{\mathcal{R}}
\DeclareMathOperator{\K}{\mathcal{K}}
\DeclareMathOperator{\Sp}{\mathbb{S}^+}

\DeclareMathOperator{\St}{\mathbb{S}}

\DeclareMathOperator{\V}{\mathcal{V}}
\DeclareMathOperator{\eq}{\underline{q}}

\DeclareMathOperator{\notR}{\backslash\hspace{-0.14em} \R}

\newcommand{\vn}{\varnothing}

\usepackage[utf8]{inputenc}
\usepackage[english]{babel}

\newcommand\ovl[1]{\overline{#1}}
\newcommand\mbf[1]{\mathbf #1}
\newcommand\avg[1]{\langle #1 \rangle}
\newcommand\mcl[1]{\mathcal #1}
\newcommand{\bigzero}{\mbox{\normalfont\Large\bfseries 0}}
\newcommand{\largeentry}[1]{\mbox{\normalfont\large $#1$}}
\newcommand{\Largeentry}[1]{\mbox{\normalfont\Large $#1$}}
\newcommand\rcp[2]{#1^+_{#2}}
\newcommand\rcm[2]{{#1}^-_{#2}}

\newcommand{\rvline}{\hspace*{-\arraycolsep}\vline\hspace*{-\arraycolsep}}

\newcommand{\tikzmark}[1]{\tikz[overlay, remember picture] \coordinate (#1);}

\begin{document}

\title{The evolution of complexity and the transition to biochemical life}

\author{Praful Gagrani$^{1}$, David Baum$^{2,3}$ }
\affiliation{$^1$Institute of Industrial Science, The University of Tokyo\\ 
$^2$Wisconsin Institute for Discovery, 
University of Wisconsin-Madison\\
$^3$Department of Botany, 
University of Wisconsin-Madison
}

\date{\today}
\begin{abstract}

While modern physics and biology satisfactorily explain the passage from the Big Bang to the formation of Earth and the first cells to present-day life, respectively, the origins of biochemical life still remain an open question. Since life, as we know it, requires extremely long genetic polymers, any answer to the question must explain how an evolving system of polymers of ever-increasing length could come about on a planet that otherwise consisted only of small molecular building blocks. In this work, we show that, under realistic constraints, an abstract polymer model can exhibit dynamics such that attractors in the polymer population space with a higher average polymer length are also more probable. 
We generalize from the model and formalize the notions of \textit{complexity} and \textit{evolution} for chemical reaction networks with multiple attractors. 
The complexity of a species is defined as the minimum number of reactions needed to produce it from a set of building blocks, which in turn is used to define a measure of complexity for an attractor. A transition between attractors is considered to be a \textit{progressive evolution} if the attractor with the higher probability also has a higher complexity. In an environment where only monomers are readily available, the attractor with a higher average polymer length is more complex. Thus, by this criterion, our abstract polymer model can exhibit progressive evolution for a range of thermodynamically plausible rate constants. We also formalize criteria for \textit{open-ended} and \textit{historically-contingent} evolution and explain the role of autocatalysis in obtaining them. Our work provides a basis for searching for prebiotically plausible scenarios in which long polymers can emerge and yield populations with even longer polymers. Additionally, the existence of features like history-dependence and open-endedness support the view that the path from chemistry to biology was one of gradual complexification rather than an instantaneous origin of life.

\end{abstract}
\maketitle

%\begingroup
%  \hypersetup{hidelinks}
%  \tableofcontents
%\endgroup

%\begin{comment}
    
\section{Introduction}

The complexity of even the simplest known genetic lifeforms suggests that a genetic apparatus could not be the result of a single phase transition during the emergence of life, but rather required a cascade of non-equilibrium phase transitions \cite{smith2016origin}. Obtaining a robust physical account of these transitions would greatly advance our understanding of the origins of life. The goal should be to develop a general description of non-equilibrium phase transition that can be applied both to the dynamics of driven, small-molecule chemistry, as might have existed prior to cellular life, and to the genetic systems that underlie Darwinian evolution today. Here we develop such a framework and explore how concepts from stochastic autocatalytic chemistry and evolutionary biology can be mapped onto one another.

Arguably the most remarkable feature of cellular life is the capacity for evolving lineages to repeatedly transition to new phases (genotypes), each of which is characterized by a distinctive set of long polymers (DNA, RNA and/or proteins). The production of specific long polymers in itself is noteworthy, since it is assumed that the environment provides monomers (or even smaller chemical building blocks) and that hydrolysis (fragmentation) reactions are likely to have higher rate constants than condensation (polymerization) reactions. This is a reasonable assumption since, otherwise, we would be in a regime in which run-away chemistry would tend to yield arbitrary mixtures of very high molecular mass compounds (so-called 'tar' or 'asphalt' \cite{benner2018mineral}). Under these assumptions, the current regime in which biology uses specific, extremely long polymers must be the result of a set of transitions to attractors enriched for longer and longer polymers. Even more remarkably, in polymer-based (genetic) evolution there seems to be no limit to the number of phases and, moreover, the phases accessible in the future tend to be highly sensitive to the current phase. Explaining these \textit{open-ended} and \textit{historically-contingent} features of biological evolution is a major challenge in origins of life research. How could simple chemistry enter a regime where long polymers are likely, where there are indefinitely many alternative phases, where viable phases can be visited in an open-ended series, and where historical contingency results in independent lineages' diverging over time?

\begin{figure}[t]
    \centering
    \includegraphics[width=.45\textwidth]{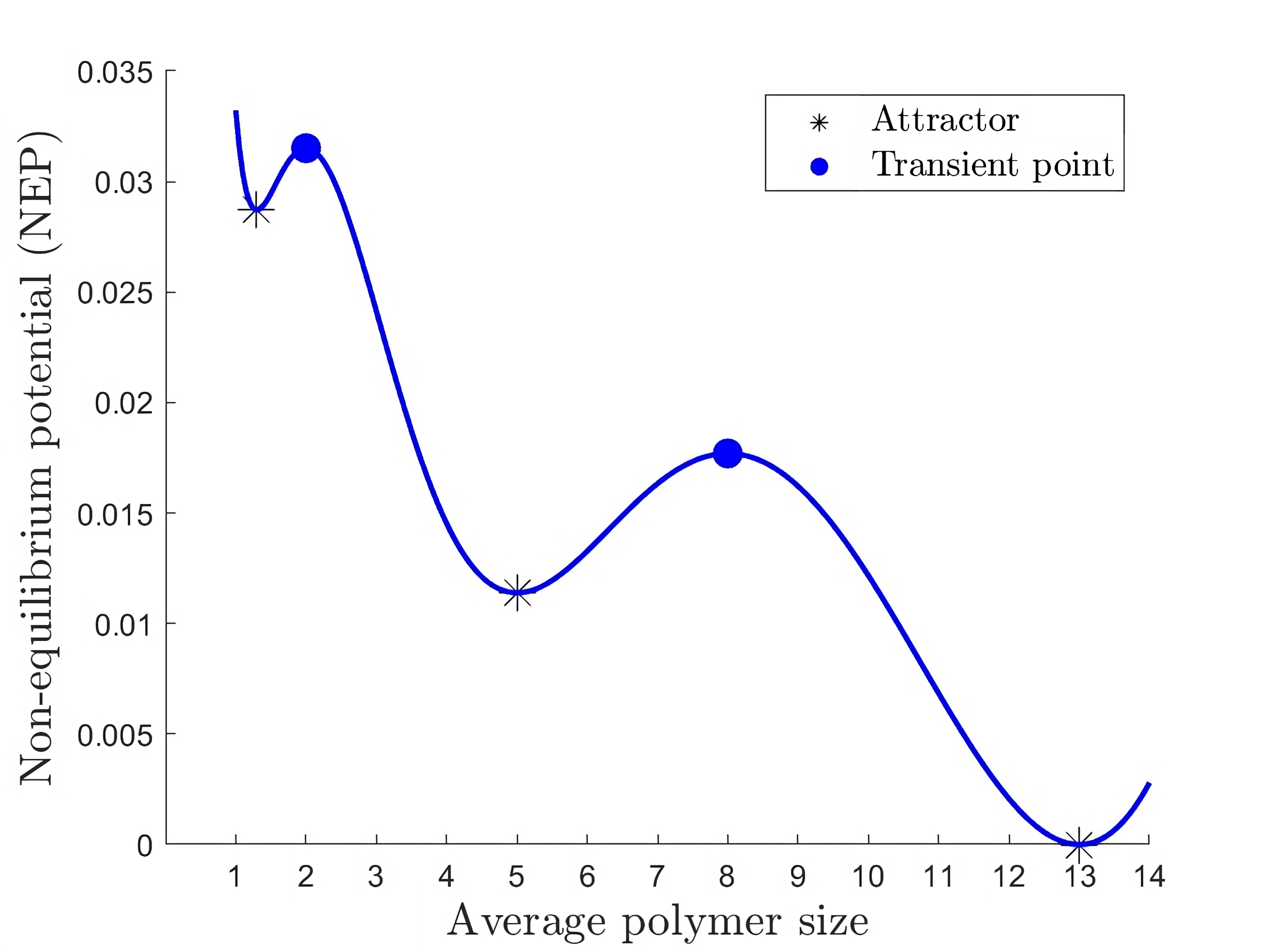}
    \caption{Example of a simple system showing progressive evolution. Three stable attractors (asterisks) and two transient points (circles) are shown, each corresponding to a population of polymers with a different average size. In this case average polymer size is a proxy for complexity. Since NEP is inversely related to stability of an attractor, and assuming that the system is driven by a flux of monomers, we would expect the system to progress over time to more complex attractors. A physically realizable model with this property (see Examples \ref{eg:1monomer_polE_model} and \ref{eg:2_1monomer_polE_model}) will imply that a system starting close to the origin will keep ratcheting itself towards attractors with higher average sizes, explaining how the preconditions to modern genetic apparatus could arise.}
    \label{fig:evo_cost}
\end{figure}

Theorists have often approached the origin-of-life problem through the concept of phase transitions, introducing models that exhibit various phases and then retrospectively categorizing them as either "life" or "nonlife" \cite{dyson1982model,mathis2017emergence,nowak2008prevolutionary,kauffman2024emergence}. In this work, we take the reverse approach by introducing a phase-transition framework for stochastic chemical reaction networks (CRNs) and mathematically defining key terms of relevance to evolutionary biology. We introduce concepts from dynamical systems, algorithmic complexity, and probability theory to provide rigorous criteria for when a stochastic CRN is capable of progressive, historically-contingent, or open-ended evolution. We construct thermodynamically realistic simple polymerization CRN models where a sequence of phases with progressively higher average polymer lengths can show monotonically increasing stability (see Fig.\ \ref{fig:evo_cost}). This, and related models developed throughout the text constitute a mathematical demonstration of conditions that permit the origins of biochemical life and a progressive origination of Darwinian evolution.

\subsection{Conceptual outline}

%Broadly, the framework proceeds as follows: starting with a stochastic system, the first step is to identify its attractors. A system with only a single attractor is incapable of evolution, making multistability a necessary prerequisite. Next, the switching dynamics of the system are numerically estimated. However, multistability alone does not suffice to qualify a system as capable of biological evolution. To address this, an additional metric of complexity is defined and evaluated at the attractors. Finally, by combining the information from the switching dynamics with the complexity of the attractors, it becomes possible to determine whether the stochastic system is capable of supporting evolutionary behavior. 

The primary contribution of this work is the development of a rigorous quantitative framework to determine whether a stochastic model can exhibit evolutionary behavior relevant to the origins of life. 
The core argument underlying our framework is that the transition from no-life to proto-life must involve a shift toward attractors of increasing complexity and greater stability. While the progression toward greater complexity may not always characterize extant life, we contend that the capacity to complexify is an essential prerequisite for the origins of life. Likewise, the state space within which life resides is assumed to be one with multiple possible attractors where stochastic transitions among attractors are structured in such a way as to result in historical contingency. Thus, we can best understand the key characteristics of life's state space by considering the set of possible attractor transitions, the probability of those transitions, and their relative complexity.

Since life is inherently chemical, this work focuses on stochastic systems that can be formalized as chemical reaction networks (CRNs) governed by stochastic chemical kinetics. The framework we develop (Sec.\ \ref{sec:math_form}) considers a complete CRN where all molecular species are included and reactions are assigned thermodynamically consistent rate constants using chemical energies of formation and reaction barrier heights. This CRN is then partitioned into internal and external subnetworks to reflect the fact that no living system can exist in isolation—it must exchange resources with its environment. Our framework accommodates an environment that is time-dependent or modeled as a CRN. In our analysis, however, a simplifying assumption is made that the environment contains a large reservoir of external species such that their concentration is assumed to be constant and independent of the state of the internal subnetwork. Using this assumption, we outline a method for numerically calculating the attractors and their stochastic switching dynamics using large-deviation theory. 

Building on this, we introduce a notion of complexity for chemical concentration profiles, derived from principles of algorithmic complexity. This allows us to associate each attractor with a measure of its complexity, which makes it possible to define criteria for different types of evolutionary behavior that a CRN can exhibit. If evolution is understood generically as a transition from one attractor to another, then a transition from a less stable and less complex attractor to a more stable and more complex attractor can be seen as \textit{progressive evolution}. Building on these considerations, we also formalize the concepts of historically contingent and open-ended evolution (Sec.\ \ref{sec:evo_criterion}), which have resisted rigorous definitions in the literature (but see for example \cite{wong2020evolutionary,packard2019overview,szathmary2006path}). Although our framework does not prescribe any specific measure of complexity, given a selected stochastic model and a chosen measure of complexity over states, our framework provides a means to ascertain whether a system is capable of progressive, historically-contingent, or open-ended evolution.          

The CRNs considered in this work include three classes of reactions: (1) reactions that transfer material out of and into the environment, (2) internal reactions within the system, and (3) reactions that couple the system to the environment, creating an autocatalytic ecology. To demonstrate our framework and key findings, we developed simplified 1- and 2-monomer polymerization CRNs. These models involve three types of reactions: (1) inflow and outflow of monomers, (2) end addition and removal of monomers from polymers, and (3) catalysis by particular polymers of monomer intake from the environment. 

Polymers that catalyze the formation of monomers, which in turn elongate other polymers, are observed in nature. For example, in modern biology, ATP synthase and kinase enzymes can activate nucleotides, the monomers of nucleic acids, by addition of phosphate moieties. We show that including the potential for catalytic polymers can cause a system to be autocatalytic. Moreover, this treatment of autocatalysis enables a mathematical simplification where the steady-state dynamics maps to the generalized Schl\"{o}gl model, allowing for closed-form analytic solutions (explicated throughout App.\ \ref{sec:math_prelim}).  

Using this simple model we demonstrate that thermodynamically feasible rate constant assignments can yield a system that exhibits progressive evolution and historical contingency, characterized by transitions toward attractors with progressively larger average polymer sizes. While these rate constants do not require fine-tuning, the behavior is not so generic that any arbitrary assignment will suffice. However, we provide a systematic approach for identifying rate constant assignments that yield progressive evolution.

To link these conclusions to general thermodynamic principles, we explore the roles of entropy production and autocatalysis using our framework and models (Sec.\ \ref{sec:remarks}). This analysis demonstrates that neither entropy production nor thermodynamic efficiency directly determine whether a system is capable of evolution. We also show that the presence of autocatalysis in a CRN is a reliable heuristic for multistability, a critical prerequisite for models aimed at explaining the origins of life.

It is possible to connect our framework with Darwinian evolutionary biology where the state of a local system (for example, a cell) is tightly correlated with the count of particular catalytic polymers, namely gene sequence variants. In Sec.\ \ref{sec:relationship_pop_gen}, we provide a conceptual map of the processes underlying the emergence of genetic life. Finally, in Sec.\ \ref{sec:discussion}, we summarize our contributions and discuss potential directions for future research.
%At this point, biologically inclined readers may turn to Sec.\ \ref{sec:evo_criterion} for definitions and examples of different types of evolution, followed by Sec.\ \ref{sec:relationship_pop_gen}. 

\section{Mathematical framework and examples}
\label{sec:math_form}

%\textcolor{blue}{Explain Fig.\ \ref{fig:dir_evo_ex}.}
In this section, we formalize the notion of \textit{evolution} for stochastic CRNs. Starting from a thermodynamically consistent CRN, we explain how to decompose it into system-environment partitions, detect how many attractors the system can have, and calculate the relative probabilities of attractors in the stationary distribution. Then, introducing tools from algorithmic complexity, we explain how to assign a measure of complexity to the different attractors, and finally provide criteria for when a CRN is capable of exhibiting evolution. As an application of the framework to the origins of biochemical life, we demonstrate a class of abstract polymer models that are capable of evolving to attractors with higher average lengths.

\begin{figure*} 
    \includegraphics[width = \textwidth]{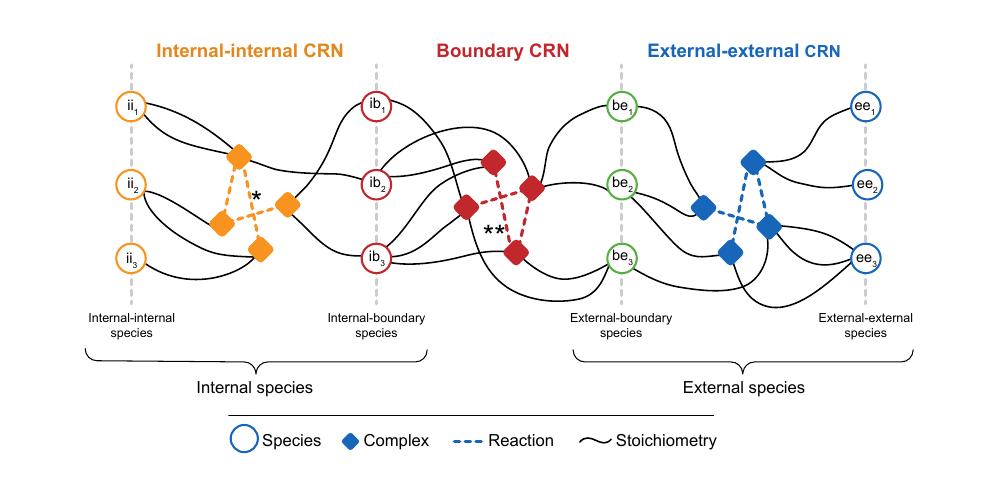}
    \caption{ Tripartite decomposition of a complete CRN into internal-internal, boundary, and external-external CRNs (Sec.\ \ref{sec:internal-external}). In the figure, following \cite{smith2017flows,smith2023rules}, we employ a representation of CRNs where the species and complexes are two types of vertices and stoichiometry and reactions are two types of edges. Irreversible reactions can be expressed using directional edges, however, here we use undirectional edges representing bidirectional reactions. For example, the starred ($*$) edge represents $\ce{2 ii_1 + ib_2 <=> ii_2 + ii_3}$, and the double starred edge ($**$) represents $\ce{ib_2 <=>[ib_3] be_3}.$}
    \label{fig:excalidraw_CRN}
\end{figure*}

\subsection{Internal-boundary-external partition}
\label{sec:internal-external}

A CRN $\mcl{G}$ taken under mass-action kinetics (MAK) will be given as a triple $$\mcl{G} = \{\Spc,\R,\K\},$$
where $\Spc,\R$ and $\K$ are the sets of species, reactions, and rate constants, respectively (the set of complexes is omitted as it can be inferred from the reaction set \cite{gopalkrishnan2011catalysis}). A brief review of CRNs, stochastic CRNs, and the thermodynamics of CRNs is given in appendices \ref{sec:CRN_intro}, \ref{sec:stoch_chemical_kin}, and \ref{sec:thermo_CRN}, respectively.  

We refer to the CRN of all the relevant species and reactions in the system and environment taken under MAK as the \textbf{complete CRN} and denote it as 
$$\mcl{G}' = \{\Spc',\R',\K'\}.$$
For \textit{thermodynamic consistency}, as explained in App.\ \ref{sec:thermo_CRN}, the rate constants should be such that the complete CRN has a detailed-balanced equilibrium. This assumption implies that, under the absence of any external control (e.g., the chemostatting of one or more external species), the concentrations of all species will relax to their equilibrium values such that the CRN produces no more entropy. As a corollary, it is implied that the CRN is reversible, and henceforth we use $\R$ to refer to the set of reactions unique up to reversal (see Eq.\ \ref{eq:reversible_MAK}).

In order to permit evolution-like behavior, we will assume that system is open, meaning that a subset of species, called \textbf{external species} and denoted by $\mcl{E}$, are controlled by the environment. The set of remaining species that relax via MAK are called \textbf{internal species} and are denoted as $\Spc$, yielding
\[ \Spc' = \Spc \cup \phantom{\cdot} \mcl{E}. \]
This partition on the species set induces an \textbf{internal-external partition} of the reactions set $\R'$ into the set of all reactions in which any internal species is consumed or produced $\R$ and its complement where no internal species participates
\[ \R' = \R \cup \notR.\]
Denoting the restriction of reactions in $\R$ to the species set $\Spc$ as $\R_{\Spc}$, we define:
\begin{align*}
    \textbf{Internal CRN  } & \mcl{G}:= \{\Spc,\R_{\Spc},\K\},\\
    \textbf{External CRN  } & \mcl{G}_e:= \{\mcl{E},\notR,\K'_e\},
\end{align*}
where $\K$ and $\K'_e$ are modified rate constants obtained from $\K'$ (see Eq.\ \ref{eq:rate_const_renorm}).  
The stoichiometric matrix of the complete CRN, denoted as $\nabla$, then takes the form 
\vspace{1em}
\begin{equation}
  \nabla = \qquad \bordermatrix{
                      ~  & \tikzmark{harrowleft} \R &  & \notR & \tikzmark{harrowright}  \cr
                    \tikzmark{varrowtop}  
                 \mbox{\normalfont\Large $\mcl{E}$}  & \mbox{\normalfont\Large $\nabla^{\R}_\mathcal{E}$} &  \begin{matrix}
  \phantom{a} \\
  \phantom{c} 
  \end{matrix} \rvline & 
  \mbox{\normalfont\Large $\nabla^{\notR}_\mathcal{E}$}\cr
  \hline \cr
                    \mbox{\normalfont\Large $\Spc$}
  & \mbox{\normalfont\Large $\St$} & \begin{matrix}
  \phantom{a} \\
  \phantom{c} 
  \end{matrix}\rvline & \bigzero \cr
  \tikzmark{varrowbottom} &  &  &  
                    },
\tikz[overlay,remember picture] {
  \draw[->] ([yshift=3ex]harrowleft) -- ([yshift=3ex]harrowright)
            node[midway,above] {\scriptsize Reactions};
  \draw[->] ([yshift=1.5ex,xshift=-2ex]varrowtop) -- ([xshift=-2ex]varrowbottom)
            node[sloped, midway,below] {\scriptsize Species};
} 
\nonumber%\label{eq:full_stoich}
\end{equation}
where $\St$ is the stoichiometric matrix of the internal CRN $\mcl{G}$.

In the absence of control, the
complete CRN would relax to the detailed-balanced equilibrium; however, in the presence
of control by the environment, the internal CRN does not necessarily do the same. Typically, the system is driven towards and maintained at a non-detailed-balanced equilibrium, a \textbf{non-equilibrium steady state (NESS)}, which is characterized by nonzero fluxes and nonzero chemical gradients \cite{qian2006open}. Moreover, in contrast to a detailed-balanced equilibrium, the \textbf{entropy production rate (EPR)} (Eq.\ \ref{eq:EPR}) at a NESS is non-zero and equals the amount of thermodynamic work done by the chemical potential energy of the external species to maintain the NESS (see App.\ \ref{sec:thermo_CRN}).

To facilitate investigation of NESSs, we further partition the internal reaction set $\R$ into internal-internal and boundary reaction sets $\R_i$ and $\R_b$, respectively. Furthermore, we partition the internal species set $\Spc$ into the \textbf{internal-boundary} set $\Spc_b$ and \textbf{internal-internal} set $\Spc_i$. Similarly, we partition the external species $\mcl{E}$ into the the external boundary, $\mcl{E}_b$, and the external-external set, $\mcl{E}_e$, yielding
\begin{align*}
   \Spc &= \Spc_b \cup \Spc_i,\\
   \mcl{E} &= \mcl{E}_b \cup \mcl{E}_e.
\end{align*}
This partition decomposes the CRN $\mcl{G}'$ into three subnetworks (as illustrated in Fig.\ \ref{fig:excalidraw_CRN}):
\begin{align*}
    \textbf{Internal-internal CRN  } & \mcl{G}'_i = \{\Spc_i \cup \Spc_b, \R_i,\K'_i\},\\
    \textbf{Boundary CRN  } & \mcl{G}'_b = \{\Spc_b \cup \phantom{\cdot} \mcl{E}_b, \R_b,\K'_b\},\\
    \textbf{External-external CRN  } & \mcl{G}'_e = \{\mcl{E}_b \cup \mcl{E}_e, \notR,\K'_e\},
\end{align*}
where $\K'_x$ is the restriction of the rate constants of $\mcl{G}'$ to $\mcl{G}'_x$ (explained below), and the stoichiometric matrix takes the form: 
\begin{equation}
  \nabla = \quad \bordermatrix{
                      ~  &  \R_i & \R_b  & & \notR   \cr            \begin{matrix}
  \largeentry{\mcl{E}_e} \\
  \largeentry{\mcl{E}_b} 
  \end{matrix}  & \bigzero  & \begin{matrix}
  \mbf{0} \\
  \nabla^{\R_b}_{\mcl{E}_b} 
  \end{matrix} &  
    \rvline &
    \Largeentry{\nabla^{\notR}_{\mcl{E}}}
    \cr
  \hline \cr
  \begin{matrix}
  \largeentry{\mcl{S}_b} \\
  \largeentry{\mcl{S}_i} 
  \end{matrix}  & \Largeentry{\St_i}  & \begin{matrix}
  \St_b \\
  \mbf{0} 
  \end{matrix} &  
    \rvline &
\bigzero \cr
                    }. \nonumber%\label{eq:stoich_w_boundary}
\end{equation}

\subparagraph{Internal-internal CRN: } 
The internal-internal CRN $\mcl{G}'_i$ consists only of reactions that do not involve any external species. As an example relevant to origins of life, $\mcl{G}'_i$ can be polymerization models \cite{andrieux2008nonequilibrium,gaspard2016kinetics}, or \textbf{cluster CRNs (CCRNs)} (Sec.\ 5, \cite{gagrani2024polyhedral}). In CCRNs, the species are counts of the constituents, forgetting bond structure, and reactions are generated by rules that satisfy certain conservation laws on the constituents. 

For example, consider polymers constituted of two monomer types $A$ and $B$ such as $AAB, ABA, AAAB,$ etc. Following notation from \cite{gagrani2024polyhedral}, a cluster species in a CCRN with $m$ constituents will be denoted as $\ovl{\mbf{n}} = \ovl{n_1,\ldots,n_m}$. Thus, in the example, both polymers $AAB$ and $ABA$ map to the cluster $\ovl{2,1}$, $AAAB$ maps to $\ovl{3,1}$, and so on (where we have arbitrarily chosen an ordering, $A,B$). The CCRN model is a coarse-graining of the polymer CRN that shows similar phenomenology but is easier to work with because the size of the species set is polynomial in a CCRN rather than exponential, as it is for polymers (see \cite{gagrani2024polyhedral} for more discussion of CCRNs). Finally, the reaction set in a CCRN is algorithmically computable and will only contain reactions that satisfy integral conservation laws, of the type
\begin{align*}
    \ce{ $\ovl{i,j} + \ovl{k,l}$ &<=> $\ovl{i\text{+}j,k\text{+}l}$},\\
    \ce{ $\ovl{i,j} + \ovl{k,l}$ &<=> $\ovl{m,n} + \ovl{i\text{+}k\text{-}m,j\text{+}l\text{-}n}$},  
\end{align*}
and so on, where $i,j,k,l, i\text{+}k\text{-}m, j\text{+}l\text{-}n \in \mathbb{Z}_{\geq 0}$.

\subparagraph{External-external CRN:} For a complete CRN $\mcl{G}'$, there are several ways to \textbf{open} the system. One could fix the rate of change in concentration of boundary species, or employ more complicated control mechanisms, such as where the external CRN controls the concentration of boundary species through feedback \cite{ranganath2024artificial}. The kind of control systems we consider here are those where the external-boundary species $\mcl{E}_b$ are \textit{chemostatted} or fixed at a constant concentration by the environment. Thus, henceforth, we assume that the external-external species set $\mcl{E}_e$ and reaction set $\notR$ are empty. 

\subparagraph{Boundary CRN:} 
 Let us assume that the external-boundary species are chemostatted at $q^*$. The rate constants $k_r \in \K_b$ on the boundary reactions $r \in \R_b$ are then obtained using $k'_r \in \K'_b$ and the chemostatted concentration such that
 \begin{equation}
     k_r = k_r' \prod_{s \in \mcl{E}_b} (q^*_s)^{r^-_s}. \label{eq:rate_const_renorm}
 \end{equation}
Notice that the re-assignment of rate constants through this procedure leads to regimes where the Wegscheider's conditions \cite{schuster1989generalization} are not satisfied leading to non-detailed balanced equilibrium. This consideration is crucial to mulstistability and we return to this point in Sec.\ \ref{sec:autocat_remarks}.

The internal-boundary and external-boundary species sets play different roles in an open-system.
As derived in Eq.\ \ref{eq:EPR_emergent}, at a NESS, the EPR is accounted for only using the chemical potential of the external-boundary species set $\mcl{E}_b$ \cite{wachtel2022free}. On the other hand, through their interaction with the external-boundary species, the internal-boundary species may break conservation laws in the internal CRN that are otherwise present in the internal-internal CRN (see E.g.\ \ref{eg:1monomer_polE_model}).

Using the above decomposition, the \textit{internal CRN} is 
\begin{align*}
   \mcl{G} &= \{ \Spc, \R,\K\} \\
    &= \{ \Spc_i \cup \Spc_b, \R_i \cup (\R_b)_i, \K'_i \cup \K_b\},
\end{align*}
and its stoichiometric matrix is given by 
\[ \St = 
\begin{pmatrix}
  \Largeentry{\St_i}  & \begin{matrix}
  \St_b \\
  \mbf{0} 
  \end{matrix} 
  \end{pmatrix}.
\]
By construction, the internal CRN can exhibit several NESSs and their entropy production is informed by thermodynamic considerations. However, as explained in Sec.\ \ref{sec:remarks_EPR_evolution}, simple thermodynamic considerations of entropy production and free-energy transduction are insufficient to formalize \textit{evolution}.

\begin{figure}
    \centering
    \includegraphics[width = 0.5\textwidth]{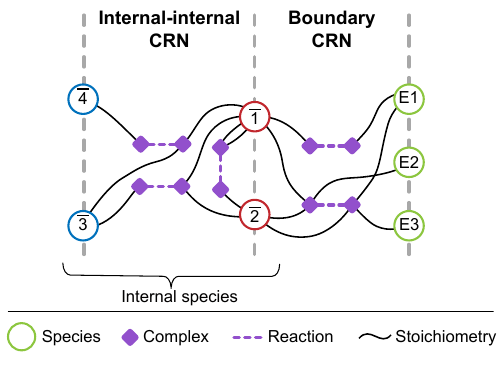}
    \caption{ Decomposition of the 1-monomer polymerization model $\mcl{P}'_{1,4}$ (E.g.\ \ref{eg:1monomer_polE_model} with $g=1$ and $N=4$). Due to chemostatting of external species, the external-external reaction set is empty and is omitted from the figure. An analysis with rate constant assignments of $\mcl{P}_{1,4}$ can be found in E.g.\ \ref{eg:2_1monomer_polE_model}.}
    \label{fig:excalidraw_CRN}
\end{figure}

\begin{example}
\label{eg:1monomer_polE_model}
Consider a class of models with relevance to the origins of biochemical life, namely \textbf{1-monomer polymerization models}. The class is parameterized by two positive integers $g$ and $N$, and we denote a complete CRN and internal CRN in the class as $\mcl{P}'_{g,N}$ and $\mcl{P}_{g,N}$, respectively. $N$ and $g$ count the number of polymers and functional species (see Sec.\ \ref{sec:autocat_remarks}), respectively, in the internal CRN.

It can be shown that the complete CRN has deficiency zero (see App.\ \ref{sec:CRN_intro} for definition and E.g.\ \ref{eg:emb_gen_schlogl_defnt} for a similar example). Since it is also reversible, by Feinberg's deficiency zero theorem (Thm.\ \ref{thm:def_zero}), the complete CRN has a unique equilibrium in each stoichiometric compatibility class for any choice of rate constants. On the other hand, the internal CRN has deficiency non-zero, and it will be shown in subsequent examples that there do exist rate constants such that the internal CRN exhibits multistability.  

The \textit{internal-internal CRN}, denoted by $(\mcl{P}'_i)_{N}$ consists of the set of reactions representing polymerization by monomer addition at an end (since it does not explicitly depend on $g$, we suppress that parameter). Since a 1-monomer polymer model is indistinguishable from a 1-constituent CCRN, we employ the cluster notation to define the CRN. 
\begin{align*}
    (\mathcal{P}'_i)_N &= \{ \Spc_N, \R_N, \K_N\}\\
    \Spc_N &= \{ \ovl{1}, \ovl{2}, \ldots, \ovl{N}\},\\
    \R_N &= \{ \ce{$\ovl{1} + \ovl{1}$ <=>[$k_{2}^+$][$k_{2}^-$] $\ovl{2}$},\\
    & \phantom{=2} \ce{$\ovl{1} + \ovl{2}$ <=>[$k_{3}^+$][$k_{3}^-$] $\ovl{3}$},\\
    & \phantom{=2} \ldots \\
    & \phantom{=2} \ce{$\ovl{1} + \ovl{N\text{-1}}$ <=>[$k_{N}^+$][$k_{N}^-$] $\ovl{N}$}\}.
\end{align*}
Employing notation from Sec.\ 5 of \cite{gagrani2024polyhedral}, we can condense the reaction set as the following rule set
\[\R_N = \{ \ce{$\ovl{1} + \ovl{j\text{-}1}$ <=>[$k_{j}^+$][$k_{j}^+$] $\ovl{j}$} \phantom{\cdot}\big| \phantom{\cdot} 1 \leq j \leq N \in \mathbb{Z}\}.\]
The stoichiometric matrix $\St_i$ for the internal-internal CRN becomes
\begin{equation*}
  \St_i = \bordermatrix{
                      ~  &   & &   &     \cr  
                      \ovl{1} &-1 & \ldots  &-1  & -2  \cr
                      \ovl{2} & \phantom{-}0 & \ldots  &-1  & \phantom{-}1  \cr
                  \ovl{3} & \phantom{-}0 & \ldots  & \phantom{-}1  & \phantom{-}0  \cr            
                  \ldots & \phantom{-}0  & \ldots  &  \phantom{-}0 & \phantom{-}0  \cr           
                  \ovl{N\text{-1}} &-1 & \ldots  & \phantom{-}0  & \phantom{-}0 \cr       
                  \ovl{N} & \phantom{-}1 & \ldots  & \phantom{-}0  & \phantom{-}0               
                    }. 
\end{equation*}

As discussed earlier, the \textit{external-external CRN} for the complete CRN will be taken to be empty. To specify the complete CRN, we now specify the \textit{boundary CRN} $(\mcl{P}'_b)_g$ (suppressing $N$ because the CRN does not explicitly depend on the parameter). The first reaction is the production of the monomer $\ovl{1}$ by the boundary species $E_0$, and all subsequent reactions are catalyzed production of the monomer in presence of different boundary species. In this model, different external boundary species have been coupled to the boundary reactions to simplify the CRN's decomposition into processes and transduction analysis (see App.\ \ref{sec:thermo_CRN}).
\begin{align*}
    (\mcl{P}'_b)_g &= \{ \Spc_g \cup \phantom{\cdot} \mcl{E}_g , (\R_b)_g, (\K_b)_g\}\\
(\Spc'_b)_g &= \{ \ovl{1}, \ovl{2}, \ovl{4}, \ldots, \ovl{2g}\},\\
(\mcl{E}_b)_g &= \{ E_0, E_2, E_3,  \ldots, E_{2g+1}\},\\
\{\mathcal{R}'_g,\mathcal{K}'_g\} &= \{\ce{E_0 <=>[${c'}_0^+$][${c'}_0^-$] \ovl{1}},\\
& \phantom{=0} \ce{E_2 + E_0 + \ovl{2} <=>[${c'}_1^+$][${c'}_1^-$] \ovl{2} + \ovl{1} + E_3},\\
 & \phantom{=0} \ldots \\
 & \phantom{=0}  \ce{E_{2g} + E_0 + \ovl{2g} <=>[${c'}_g^+$][${c'}_g^-$] \ovl{2g} + \ovl{1} + E_{2g+1}}\}.
\end{align*}
The stoichiometric matrix for the boundary CRN is
\begin{equation*}
   \bordermatrix{
      ~  &   & &   &     \cr  
      E_{2g+1} & \phantom{-} 1 & \ldots  & \phantom{-}0  &  \phantom{-}0  \cr
      E_{2g} & -1 & \ldots  & \phantom{-}0  & \phantom{-}0  \cr
      \ldots & \phantom{-}0  & \ldots  &  \phantom{-}0 & \phantom{-}0  \cr
      E_3 & \phantom{-} 0 & \ldots  &\phantom{-}1  & \phantom{-}0  \cr
      E_2 & \phantom{-}0 & \ldots  &-1  & \phantom{-}0  \cr
  E_0 & -1 & \ldots  & -1  & -1  \cr 
  \hline \cr
  \ovl{1} & \phantom{-}1  & \ldots  &  \phantom{-}1 & \phantom{-}1  \cr      
  \ovl{2} & \phantom{-} 0 & \ldots  & \phantom{-}(1)  & \phantom{-}0 \cr       
  \ldots & \phantom{-}0  & \ldots  &  \phantom{-}0 & \phantom{-}0  \cr
  \ovl{2g} & \phantom{-} (1) & \ldots  & \phantom{-}0  & \phantom{-}0        
                    },
\end{equation*}
yielding the stoichiometric matrix of the external-boundary motif
\begin{equation*}
  \nabla^{\mcl{E}}_{\Spc} = \quad \bordermatrix{
      ~  &   & &   &     \cr  
      E_{2g+1} & \phantom{-} 1 & \ldots  & \phantom{-}0  &  \phantom{-}0  \cr
      E_{2g} & -1 & \ldots  & \phantom{-}0  & \phantom{-}0  \cr
      \ldots & \phantom{-}0  & \ldots  &  \phantom{-}0 & \phantom{-}0  \cr
      E_3 & \phantom{-} 0 & \ldots  &\phantom{-}1  & \phantom{-}0  \cr
      E_2 & \phantom{-}0 & \ldots  &-1  & \phantom{-}0  \cr
  E_0 & -1 & \ldots  & -1  & -1  \cr 
                    }
\end{equation*}
and the stoichiometric matrix of the internal-boundary CRN 
\begin{equation*}
  \St_b = \quad \bordermatrix{
      ~  &   & &   &     \cr  
  \ovl{1} & \phantom{-}1  & \ldots  &  \phantom{-}1 & \phantom{-}1  \cr      
  \ovl{2} & \phantom{-} 0 & \ldots  & \phantom{-}(1)  & \phantom{-}0 \cr       
  \ldots & \phantom{-}0  & \ldots  &  \phantom{-}0 & \phantom{-}0  \cr
  \ovl{2g} & \phantom{-} (1) & \ldots  & \phantom{-}0  & \phantom{-}0        
                    }. 
\end{equation*}
Since the reactions involve direct catalysis (same species in reactant and product sets) but the stoichiometric matrix misses that information, we denote the stoichiometric coefficient of the catalyzed species in parentheses. 

Finally, we chemostat the boundary species $\{E_0,E_2, \ldots,E_{2g+1}\}$ to obtain the reactions and rate constants (Eq.\ \ref{eq:rate_const_renorm}) of the internal CRN $\mcl{P}_{g,N}$. The reaction set of the internal CRN consists of the polymerization reactions, a single monomer production reaction, and catalyzed monomer production reactions. Assuming $N\geq 2g$, the internal CRN is given as:
\begin{align*}
    \mcl{P}_{g,N} &= \{ \Spc_{N,g}  , \R_{N,g}\}\\
       \Spc_{N,g} &= \{ \ovl{1}, \ovl{2}, \ldots, \ovl{N}\},\\
(\R_b)_g &= \{\ce{$\vn$ <=>[$c_0^+$][$c_0^-$] \ovl{1}},\\
& \phantom{=0} \ce{\ovl{2} <=>[$c_1^+$][$c_1^-$] \ovl{2} + \ovl{1} },\\
 & \phantom{=0} \ldots \\
 & \phantom{=0}  \ce{\ovl{2g} <=>[$c_g^+$][$c_0^-$] \ovl{2g} + \ovl{1}}\}\\
 \R_{N,g} &= \R_N \cup (\R_b)_g.
\end{align*}
\end{example}

\subsection{Attractor-transition graph}
\label{sec:attractor_transition}

Following terminology from dynamical systems theory, we identify NESSs, or \textit{attractors}, of an internal CRN with stable fixed points of the mass-action kinetics of the mean values of populations. In this subsection, we explain how to determine the attractors of a stochastic CRN and the expected time required by the system to escape each attractor. We can summarize these results in an \textbf{Attractor-Transition (A-T) graph}, defined as follows:
\begin{itemize}
    \item Any two attractors adjacent to the same \textit{transient point} are connected in the A-T graph.
    \item The direction points towards the attractor with a higher probability in the stationary distribution.
\end{itemize}
We use the A-T graph to formalize a concept of evolution in Sec.\ \ref{sec:evo_criterion}.

The relationship between a Markov process specified by a local transition-rate matrix to conditional probabilities of events separated in state space and time is well-understood through path-integral methods (see Ch.\ 10.4 in \cite{altland2010condensed} and Sec.\ VI 3 in \cite{touchette2009large}). A short introduction and reference to mathematical and physics literature can be found in App.\ \ref{sec:stoch_chemical_kin}. A brief summary of the results needed for the construction of A-T graph is as follows.  

In a stochastic CRN, \textit{events} are specified by the vector of species \textbf{population} and \textbf{time}
\[ (n,t) \text{ for } n \in \mathbb{Z}^{\Spc}_{\geq 0}, t \in \mathbb{R},\]
along with a \textbf{volume} (also called \textit{scale factor}) $V$ of the system which yields the species \textbf{concentration} 
\begin{equation}
q = \frac{n}{V} \in \mathbb{Q}^{\Spc}_{\geq 0}. \label{eq:conc_population}    
\end{equation}
For a reversible CRN, the \textbf{Hamiltonian} is given as 
\begin{equation}
      H(p,q) = \sum_{r \in \mathcal{R}} 
      (e^{r^+\cdot p}-e^{r^-\cdot p})\left( k_{r}^+(e^{-p}q)^{r^-} 
      -k_{r}^-(e^{-p}q)^{r^+} \right), \label{eq:reversible_Hamiltonian}
\end{equation}
where $p \in \mathbb{R}^{\Spc}$ is the \textit{momentum} vector canonically conjugate to the concentrations.
As explained in App.\ \ref{sec:stoch_chemical_kin}, the Hamiltonian function can be used to obtain leading order approximations to conditional probabilities of events, and through that, other quantities for understanding the stochastic dynamics of the system.

Consider the Hamiltonian from Eq.\ \ref{eq:reversible_Hamiltonian} for an internal CRN. As explained in App.\ \ref{sec:stoch_chemical_kin}, the equations of MAK are obtained as (Eq.\ \ref{eq:relax_Ham})
\begin{equation}
    \dot{q} = \pdv{H}{p}\bigg|_{p=0}. \label{eq:MAK_Ham_sec}
\end{equation}
The set of all concentrations $\eq$ such that 
\[ \pdv{H}{p}\bigg|_{(q,p) = (\eq,0)} = 0\]
are the \textit{fixed points} of the MAK. If we determine the eigenvalues of the Hessian 
\[ J = \pdv{H}{p}{q}\bigg|_{(q,p) = (\eq,0)}\]
at the fixed points, an \textbf{attractor} or a \textbf{transient point} are, respectively, fixed points with zero or a single eigenvalue with a strictly positive real part. As mentioned above, we define a NESS as an attractor of the internal CRN and use the two interchangeably.

The solutions of the mass-action equations in Eq.\ \ref{eq:MAK_Ham_sec} are called \textit{decay} or \textbf{relaxation} trajectories and connect states $q$ to fixed points $\eq$. A transient point is \textbf{adjacent} to an attractor if there is a mass-action solution that starts in the vicinity of the transient point and terminates at the attractor. 
In stochastic CRNs, one can also find \textit{fluctuation} or \textbf{escape} trajectories that take the system out of a NESS $\eq$ to a state $q$. As explained around Eq.\ \ref{eq:NEP_diff}, the momentum assignment along an escape trajectory is used to obtain the difference in \textbf{non-equilibrium potential} (NEP), denoted as $\V(q)$, which in turn can be used to obtain the stationary distribution $\pi(n)$ (when it exists,  \cite{anderson2010product,hoessly2021stationary}) of the stochastic process using
\[ \pi(n=Vq) \asymp e^{-V \V(q)}.\]
In \cite{gagrani2023action}, we give a computational algorithm to estimate the NEP for multistable CRNs by finding the most probable escape or fluctuation paths.

The proportion of time the system spends in state $q$ converges to its stationary distribution value $\pi(q)$ as time runs to infinity (Ch.\ 1, \cite{kelly2011reversibility}). The NEP can also be used to calculate the mean \textbf{escape time} of an attractor out of its domain of attraction. 
Let the set $t^a = \{t^a_1,\ldots,t^a_N\}$ be the set of transient points adjacent to an attractor $a$. Let the NEP difference between $t^a_i$ and $a$ be given be $\Delta \V_i(a) \geq 0$. Define $\Delta \V^*(a)$ to be the minimum of $\Delta \V_i(a)$
\[\Delta \V^*(a) = \min_i \Delta \V(a).\]
Then, from Eq.\ 240 \cite{touchette2009large}, the mean \textit{escape time} from the domain of attraction of $a$ is given by 
\begin{equation}
    \tau_\text{esc}(a) = e^{V \Delta \V^*(a)}. \label{eq:escape}
\end{equation}
We can see the time spent by a system in an attractor gives a measure of the \textbf{stability} of the attractor. Thus, the depth of the NEP well around an attractor is a direct measure of the stability of the attractor. 

We conclude this section by describing a strategy for determining how the depth of the NEP wells around an attractor can be adjusted by varying the rate constants. Our argument relies on the continuity of the map from rate constants to fixed points of the MAK. 
It is well-known (rederived in Th.\ \ref{thm:Lyapunov}) that the NEP is a \textit{Lyapunov function} for solutions of MAK. Suppose a transient point $t$ has two adjacent attractors $a_1^t$ and $a_2^t$. Thus, the NEP at attractors is lesser or equal than the value at the adjacent transient point
\[ \V(t) \geq \V(a_1^t), \hspace{0.2em} \V(t) \geq \V(a_2^t).\]
This simply means that in the stationary distribution, the stochastic system is less likely to be at the transient point than the attractors.
However, notice that if the transient point $t$ was identified with an attractor, say $a_1$, i.e., $t = a_1$, then 
\[ \V(t) = \V(a_1)> \V(a_2).\]
This means that $a_2$ would have a higher probability in the stationary distribution (lower NEP). By the same argument, if the transient point was very close to $a_2$, the probability in the stationary distribution of $a_1$ will be more than that of $a_2$. Using this observation, at least in simple cases, we can mathematically \textit{engineer} the rate constants of multistable CRNs with a desired ordering on the stationary distribution. (Also, see E.g.\ \ref{eg:ex4_3models_gen_schlogl}).

\begin{figure*}
    \centering
    \begin{subfigure}[b]{0.475\textwidth}
        \centering
        \includegraphics[width=\textwidth]{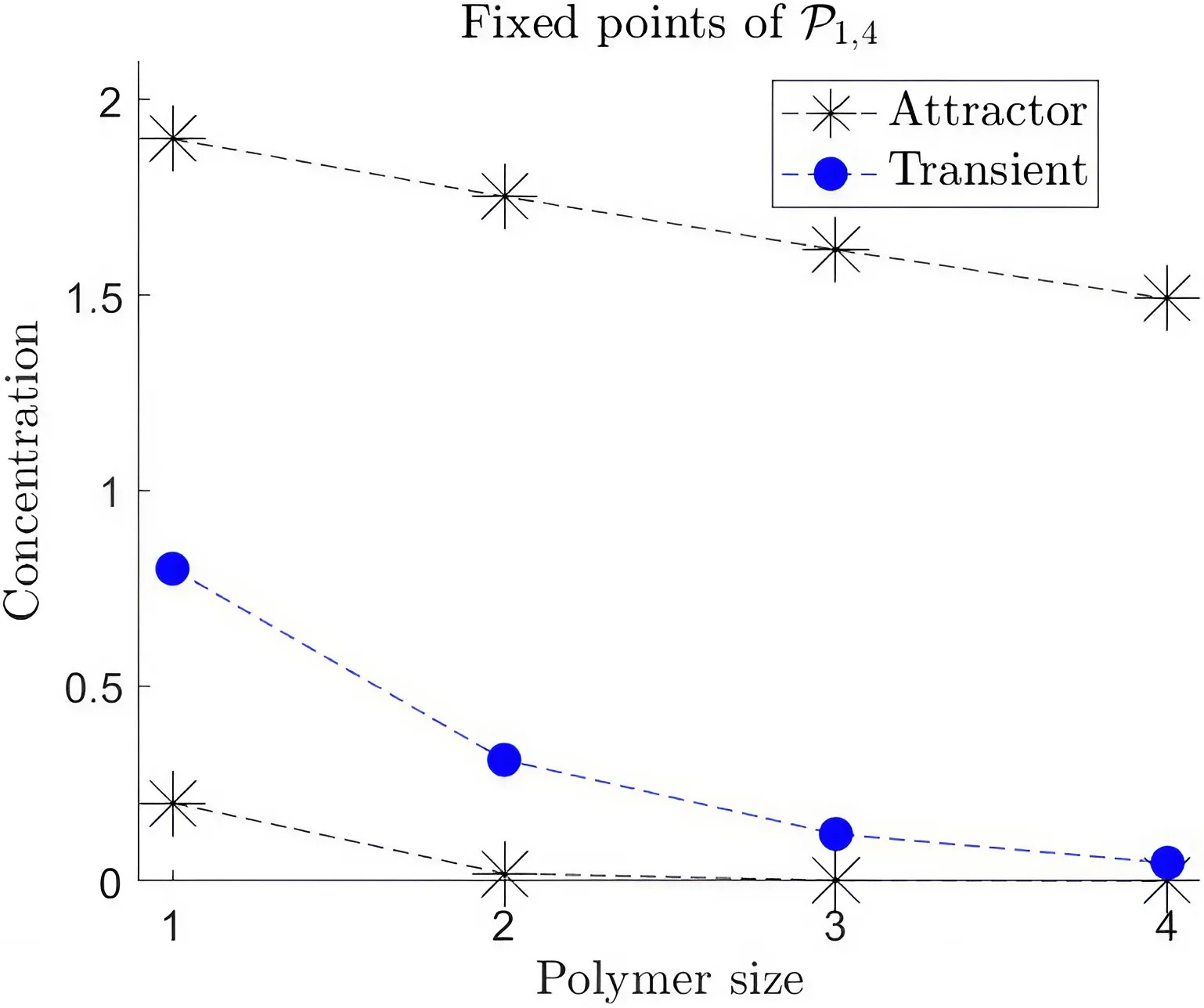}
        \caption[]{}
        %{{\small Stochastic simulation}}    
        %\label{fig:mean and std of net14}
    \end{subfigure}
    \hfill
    \begin{subfigure}[b]{0.475\textwidth}  
        \centering 
        \includegraphics[width=\textwidth]{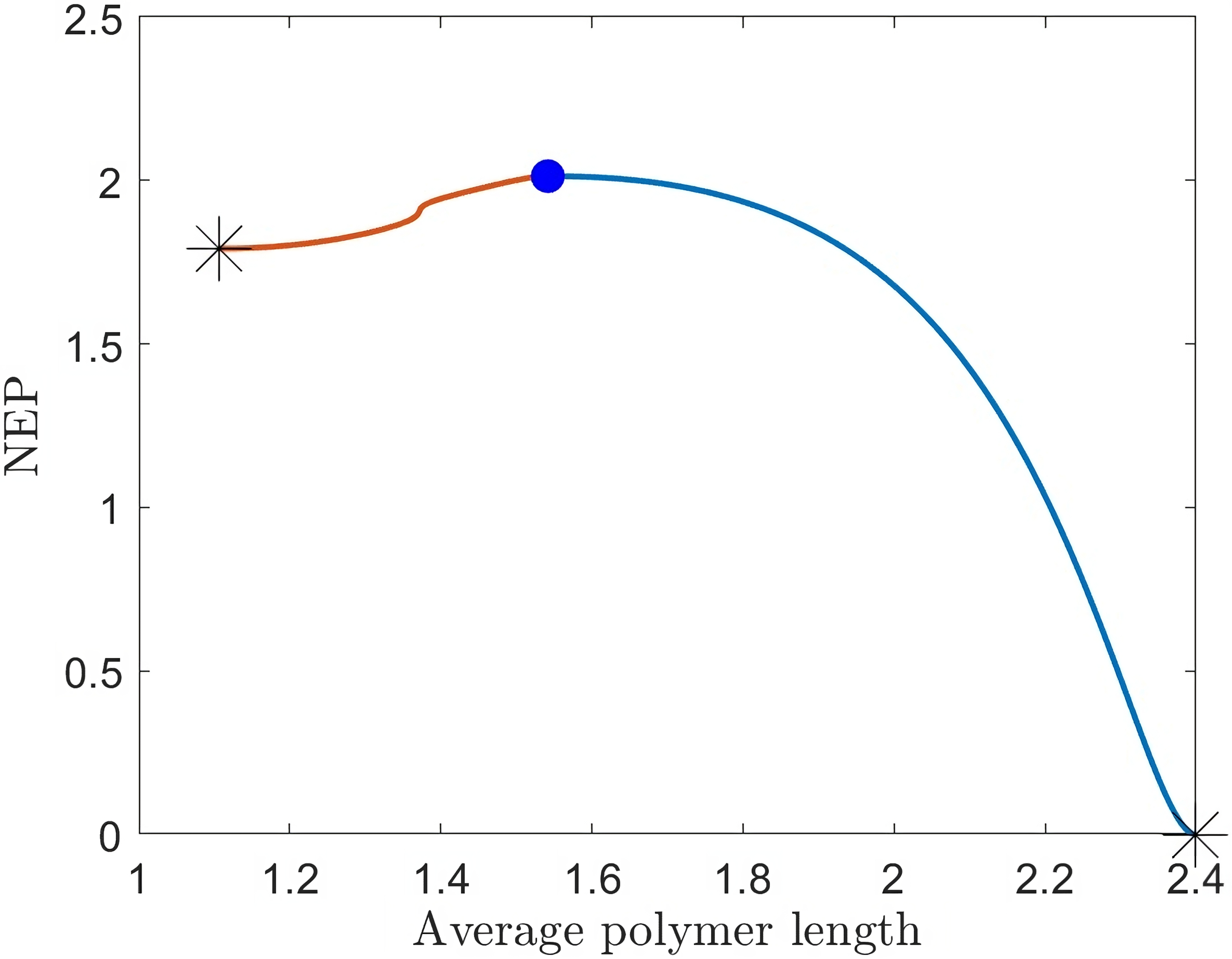}
        \caption[]{}
        %{{\small Model 1}}    
        %\label{fig:mean and std of net24}
    \end{subfigure}
    \vskip\baselineskip
    \begin{subfigure}[b]{0.475\textwidth}   
        \centering 
        \includegraphics[width=\textwidth]{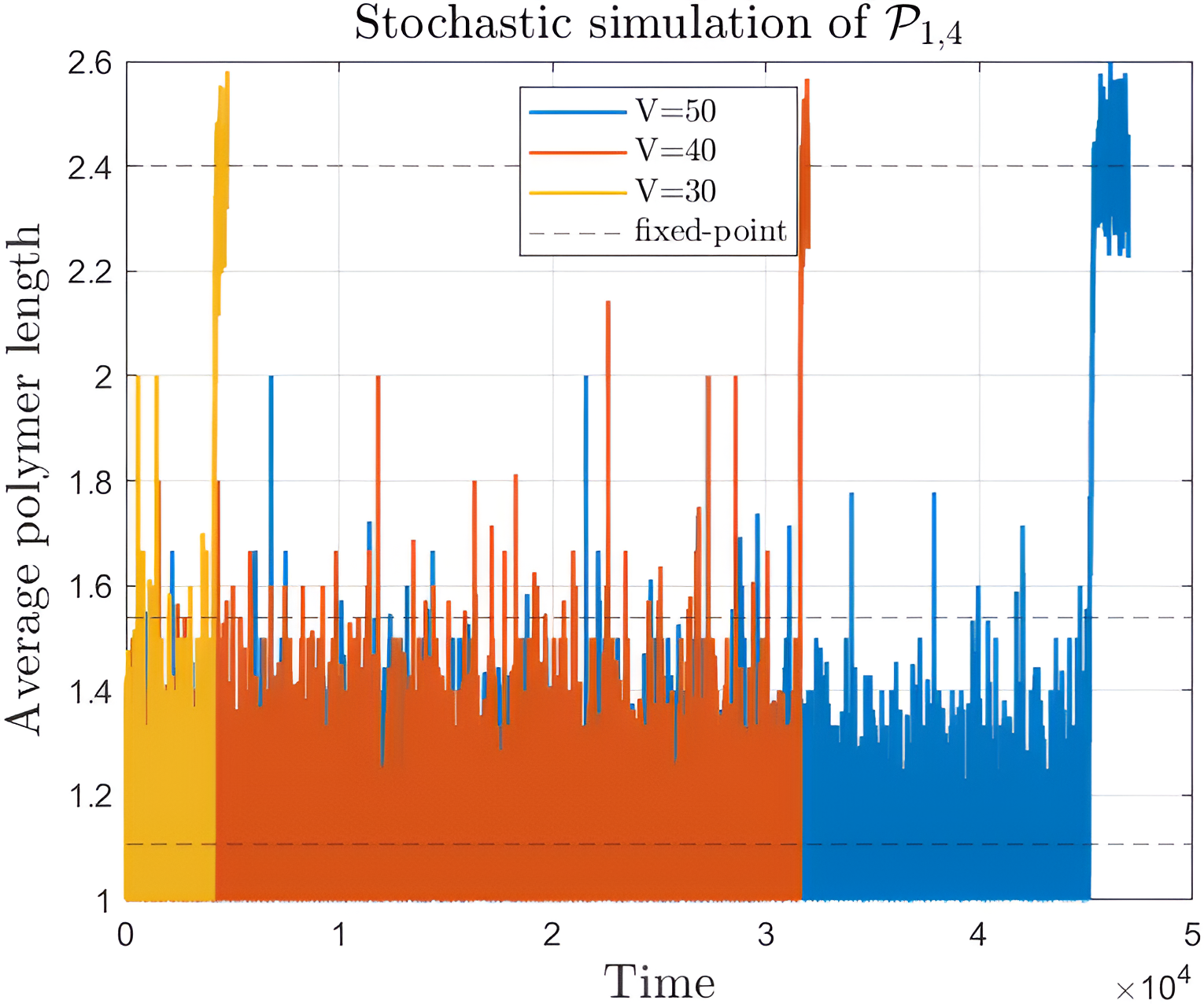}
        \caption[]{}    
        %\label{fig:mean and std of net34}
    \end{subfigure}
    \hfill
    \begin{subfigure}[b]{0.475\textwidth}   
        \centering 
        \includegraphics[width=\textwidth]{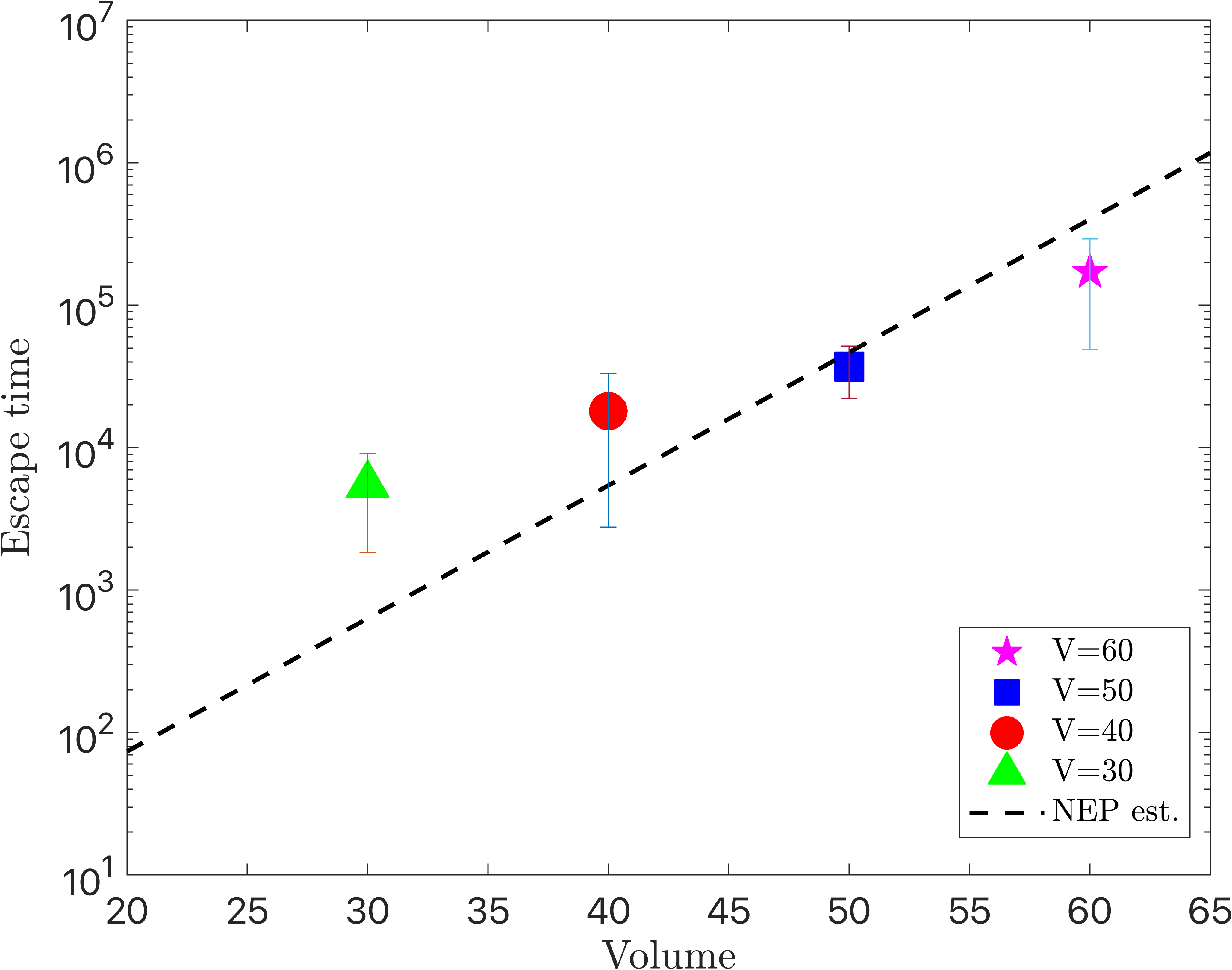}
        \caption[]{}
        %{{\small Model 3}}    
        %\label{fig:mean and std of net44}
    \end{subfigure}
    \caption[Aspects of the internal CRN from Example \ref{eg:2_1monomer_polE_model}]
    { Worked example showing the fixed-points, NEP, stochastic simulation results, and escape times of the internal CRN from example \ref{eg:2_1monomer_polE_model}.
 (a) Distributions of polymer length at the three fixed points. (b) An estimate for the NEP along the escape paths linking the two attractors parameterized by the average polymer length, as obtained using the AFGD algorithm \cite{gagrani2023action}. (c) Sample stochastic simulations for different volumes illustrating that smaller systems transition more rapidly between adjacent attractors than larger systems. Simulations are only shown through the first time the higher attractor is visited. (d) Escape times from 10 stochastic simulations overlaid on a numerical estimate using the value of the NEP at the attractor with the lower average polymer length. 
\label{fig:1monomer_ex2}} 
\end{figure*}

\begin{figure*}[]
     \centering
  \includegraphics[width=0.93\linewidth]{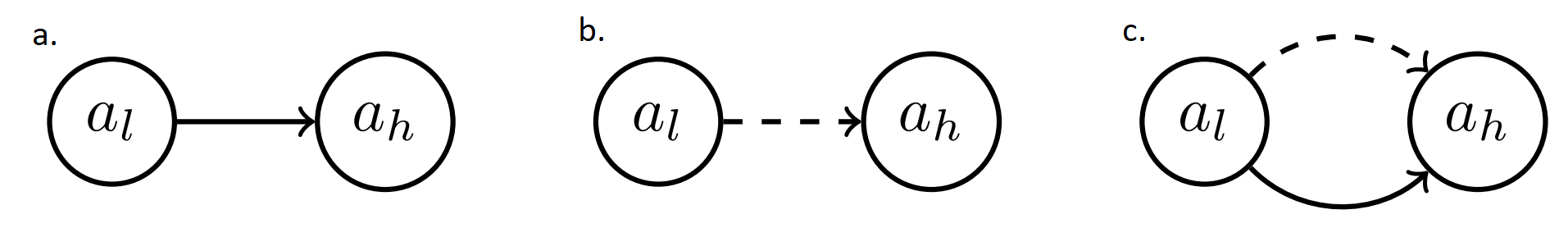}
  \caption{An example of progressive evolution. The $a.$ attractor-transition graph, $b.$ complexity graph, and $c.$ evolution graph for the 1-monomer polymerization model $\mcl{P}_{1,4}$ from E.g.\ \ref{eg:2_1monomer_polE_model}. Since attractor $a_h$ is both more stable and more complex than $a_l$, the system will be said to be capable of progressive evolution from $a_l$ (Sec.\ \ref{sec:evo_criterion}).}
  \label{fig:dir_evo_ex}
\end{figure*}

\begin{example}
\label{eg:2_1monomer_polE_model}
Consider the class of internal CRNs denoted as $\mcl{P}_{g,N}$ from E.g.\ \ref{eg:1monomer_polE_model}.
While the Hamiltonian for this model can be explicitly written, it is not particularly enlightening. Instead, we directly write the equations of MAK. Denoting the concentration of species $\overline{i}$ by $q_i$, and its rate of change by $\dot{q}_i$, the equations of MAK are
\begin{align*}
    \dot{q}_n &= - \rcm{k}{n}q_n + \rcp{k}{n}q_1q_{n-1}, \\
    \dot{q}_{n-1} &=  \rcm{k}{n}q_n - \rcp{k}{n}q_1q_{n-1}- \rcm{k}{n-1}q_{n-1} + \rcp{k}{n-1}q_1q_{n-2},\\
    &\ldots \\
    \dot{q}_{2} &=  \rcm{k}{3}q_3 - \rcp{k}{3}q_1q_{2} - \rcm{k}{2}q_{2} + \rcp{k}{2}q_1^2,\\
    \dot{q}_1 &= ( \rcm{k}{2}q_{2} - \rcp{k}{2}q_1^2) + (\rcp{c}{0} - \rcm{c}{0}q_1)\\
   & \phantom{=} + \sum_{i=2}^{N} \left(\rcm{k}{i}q_i - \rcp{k}{i}q_1 q_{i-1}\right) \\
    & \phantom{=} +\sum_{j=1}^g\left( \rcp{c}{j} - \rcm{c}{j}q_1\right)q_{2j}. 
\end{align*}

The fixed points of the system are obtained by solving for $\eq$ such that $\dot{q}(\eq) = 0$. Elementary considerations yield that the fixed points satisfy
\begin{align*}
    \underline{q_i} &= \prod_{j=2}^{i}\left[\frac{\rcp{k}{j}}{\rcm{k}{j}}\right] (\underline{q_1})^i \text{ for } i \in [2,N],\\
    0 &= (\rcp{c}{0} - \rcm{c}{0}\underline{q_1})+\sum_{j=1}^g\left( \rcp{c}{j} - \rcm{c}{j}\underline{q_1}\right)\underline{q_{2j}}.
\end{align*}
Substituting $\underline{q_{2j}}$ from the first equation into the second equation yields a polynomial whose roots are the equilibrium concentrations of the monomer. A simple application of Descartes' rule of signs yields that the maximum number of positive real roots is $2g+1$. Thus, we conclude that $\mcl{P}_{g,N}$ can have up to $g+1$ attractors.

As a concrete example, consider an internal CRN with $g=1$ and $N=4$ denoted as $\mcl{P}_{1,4}$
    \begin{align*}
    \mcl{P}_{1,4} &= \{ \Spc_{4,1}  , \R_{4,1}\}\\
       \Spc_{4,1} &= \{ \ovl{1}, \ovl{2}, \ovl{3}, \ovl{4}\},\\
        \R_{4,1} &= \{\ce{$\vn$ <=>[$c_0^+$][$c_0^-$] \ovl{1}},\\
& \phantom{=0} \ce{2\ovl{1} <=>[$k_2^+$][$k_2^-$] \ovl{2} <=>[$c_1^+$][$c_1^-$] \ovl{2} + \ovl{1} <=>[$k_3^+$][$k_3^-$] \ovl{3} },\\
 & \phantom{=0}  \ce{\ovl{1} + \ovl{3} <=>[$k_4^+$][$k_4^-$] \ovl{4}} \},
\end{align*}
with rate constant assignment
\begin{align*}
    k^+ &= 1 = \rcp{k}{i} &\text{ for } i &\in [2,4],\\
    k^- &= 2.06 = \rcp{k}{i} &\text{ for } i &\in [2,4],\\
    \rcp{c}{0} &= 0.1476, & \rcm{c}{0} &= 1,\\
    \rcp{c}{1} &= 2.9, & \rcm{c}{1} &= 1.
\end{align*}
The vectors of fixed points $\eq = [\underline{q_1},\underline{q_2},\underline{q_3},\underline{q_4}]$ for this model are 
\begin{align*}
    \eq_1 &= [ 0.20, 0.019, 0.009, 0.00018]\\
    \eq_2 &= [ 0.80, 0.31, 0.12, 0.046]\\
    \eq_3 &= [ 1.9, 1.7, 1.6, 1.5].
\end{align*}
The first and the third fixed points are attractors and the second fixed point is a transient point, as shown in Fig.\ \ref{fig:1monomer_ex2}, panel (a). 
Defining the average polymer length at a concentration as 
\begin{equation}
    \avg{\ell}_q = \sum_{i=1}^N i \frac{q_i}{\sum q}, \label{eq:avg_ell}
\end{equation}
the average polymer lengths at the three fixed points are:
\[ \langle \ell \rangle_1 = 1.1072, \hspace{1em} \langle \ell \rangle_2 = 1.54,\hspace{1em}  \langle \ell \rangle_3 = 2.399. \]
In the remainder of the example, we will call the first and third attractor the \textit{lower} and \textit{higher} attractor, respectively.

The escape paths joining the attractors to the transient point and the NEP along them are estimated using the \textit{action-functional gradient descent} (AFGD) algorithm \cite{gagrani2023action} and shown Fig.\ \ref{fig:1monomer_ex2}, panel (b), with average polymer length on the \textit{x-axis}. As will be justified in the succeeding paragraph, we do not descend to the escape path completely but only run the algorithm for a few iterations to obtain an estimate. Thus, what is shown is an upper bound to the NEP. It can be seen that the attractor with the higher average polymer length (higher attractor) has a lower NEP value than that of the lower attractor. This means that if we run a stochastic simulation for long enough, the process will spend more time in the vicinity of the higher attractor than the lower one. The A-T graph for this model is drawn in panel $a$ of Fig.\ \ref{fig:dir_evo_ex}.

Stochastic simulations for different volumes or scale factors, namely $30$, $40$, and $50$, using the Gillespie algorithm \cite{gillespie2007stochastic} are displayed in Fig.\ \ref{fig:1monomer_ex2}, panel (c). The escape time data from 10 stochastic simulations is shown in the bottom-right panel of the same figure, alongside a numerical estimate derived by calculating the NEP difference between the lower attractor and the transient point and substituting it into Eq.~\ref{eq:escape}. The numerical value of the NEP difference used here is $0.215$. Observe that the escape time is presented on a logarithmic scale. %Thus, we can conclude that the estimated prediction fits the observations rather well.

Theoretically, the NEP provides a good approximation of the stochastic simulation data in the large-volume limit. For smaller volumes—i.e., when the total number of particles is low—the population often goes extinct, and it takes relatively longer to escape the basin of attraction of the lower attractor. On the other hand, for larger volumes, the theoretical prediction overshoots the simulation data. This is because the estimate provides an upper bound for the true NEP, which can be determined with higher numerical precision using the AFGD algorithm.

Stochastic simulation data for escapes from the higher attractor are omitted because the difference in value of the NEP at the transient point and higher attractor is a factor of $10$ higher, which means that the time of escape from the higher attractor is $e^{10}$ ($\approx 22,000$) more than that at the lower attractor. This means that once the system transitions to the higher attractor, it is exponentially unlikely for it to switch back stochastically.
\end{example}

\subsection{Complexity graph}

\label{sec:complexity}

In the previous subsection, we formalized NESSs as attractors of the internal CRN and outlined how to obtain an A-T graph. In this subsection, we define a \textit{complexity function} $\mathfrak{C}$ for attractors
\begin{equation}
    \mathfrak{C}: \text{NESS} \times \text{Complexity} \to \mathbb{R}_{\geq 0} \label{eq:complexitE_NESS}
\end{equation}
and define the \textbf{complexity graph} as follows. For every pair of adjacent attractors $a_1$ and $a_2$ in the A-T graph, make a directed edge $a_1 \to a_2$ if $\Cpx(a_1) < \Cpx(a_2)$. Thus, in the directed complexity graph, the arrow points towards the attractor with the higher complexity. 

\subsubsection{Complexity of species}
In computer science and mathematics, the subfield of \textit{algorithmic complexity} assigns complexity measures to strings of data based on the complexity of algorithms that generate them  \cite{zenil2020methods}. A popular measure is the \textbf{logical depth} introduced by Bennett \cite{bennett1988logical}, which is defined as follows. Denote some universal computer as $U$, a program as $p$, the output of the computer on the program as $U(p)$, and the number of computational steps needed to perform $p$ as $T(p)$. The logical depth $D$ of a string $x$ is defined as the least number of computational steps needed to obtain it, or formally
\begin{equation}
   D(x) = \min \{ T(p) | U(p) = x\}. \label{eq:logical_depth}
\end{equation}

Logical depth has been extended to physical systems as \textit{thermodynamic depth} \cite{lloyd1988complexity}, which quantifies self-organization by measuring the reduction in entropy of the constituent parts. Prior work has applied these principles to consider the concept of complexity in chemistry (see \cite{krzyzanowski2023spacial,bajusz2015tanimoto,uthamacumaran2024salient} and references therein). It is important to note, however, that our framework is independent of the specific choice of complexity measure.

 In mathematical and computational chemistry, it is convenient to model molecules as undirected graphs. The mechanism by which a molecule is assembled can be understood in multiple ways, each leading to different measures of complexity. Here, we propose one method for assigning complexity and discuss another approach, namely \textit{assembly theory}, in Sec.\ \ref{sec:Assembly_theory}.
 
 First, chemical reactions can be considered as \textit{graph rewriting operations} drawn from a \textit{graph grammar} corresponding to reaction mechanisms \cite{mann2013graph,andersen2013inferring,andersenadefining}. The species and reaction sets of a CRN can then be seen as generated through repeated application of graph-rewriting operations on a starting set of molecules. Analogously to the logical depth for strings, we can define a measure of complexity for a molecular species as the minimum number of rewriting operations needed to produce that molecule.

In order to assign a complexity measure to a molecular species,  we provide a \textit{network expansion algorithm} \cite{ebenhoh2004structural,handorf2005expanding}. Let us assume that we are given a complete CRN generated
by an underlying set of graph-rewriting rules along
with a set of external species readily available from the environment. Define a sequence of sets $S_0, S_1, \ldots, S_N$ such that:\\
\textbf{Step $\mbf{0}$:} Introduce all external species in set $S_0$
\[ S_0 = \{ s | s \in \mcl{E}\}\]
\textbf{Step $\mbf{I}$:} Construct the set $S_I$ by including all species from the previous set $S_{I-1}$ and adding any species that can be produced by reactions consuming only species in $S_{I-1}$
\[ S_I = \{ s | s \in S_{I-1} \cup \{\text{supp}(r^+)|\text{supp}(r^-) \in S_{I-1} \hspace{0.2em} \forall r \in \R \}\}.\]
\textbf{Termination:} Stop the algorithm at iteration $N$ when the sets converge, i.e.,
\[S_N=S_{N+1}.\]
We define the \textbf{complexity of species} $C(s)$ of species $s$ as (compare with Eq.\ \ref{eq:logical_depth})
\begin{equation}
    C(s) = \begin{cases}
        \min \{ i | s \in S_i \},\\
        \infty \text{ if } s \not\in S_N.
    \end{cases}. \label{eq:complexity}
\end{equation}

\subsubsection{Complexity of a concentration profile}
We began by making a choice of a measure of complexity for species
\begin{equation}
    C: \text{Species} \to \mathbf{R}_{\geq 0}, \label{eq:complexity_abstract}
\end{equation}
for Eq.\ \ref{eq:complexity} (or \textit{assembly index}).  Now we use the complexity of species to induce a measure of complexity $\Cpx(q)$ for a concentration 
\begin{equation}
    \Cpx:  \mathbb{R}^{\Spc}_{\geq 0} \times C \to \mathbf{R}_{\geq 0}. \label{eq:complexitE_concentration}
\end{equation}
Since a NESS is specified by a concentration profile over species, we use this measure to assign a measure of complexity to attractors (Eq.\ \ref{eq:complexitE_NESS}) and, thereby, obtain the complexity-graph described in the introduction to this subsection. Through any choice of non-linear function $h$ which depends on the concentration and complexity of species, one can abstractly define a \textbf{complexity of concentrations} as
\begin{equation}
    \Cpx(q,C) = \sum_s h(q,C(s)). \label{eq:complexity_h}
\end{equation}
For this work, we investigate two of these choices, namely average and total complexity. \\
\textbf{Average complexity:} sum of species complexity weighted by their relative concentrations
\begin{equation}
\Cpx^A(q,C) = \sum_s \frac{q_s}{\sum_t q_t} C(s).  \label{eq:avg_comp}   
\end{equation}

\textbf{Total complexity:} sum of species complexity weighted by their absolute concentrations 
\begin{equation}
    \Cpx^T(q,C) = \sum_s q_s C(s).  \label{eq:total_comp}
\end{equation}

\subsubsection{Assembly theory}
\label{sec:Assembly_theory}

The complexity $C(s)$, as defined in Eq.\ \ref{eq:complexity}, is influenced by both the underlying rule-generated network and the initial set $S_0$ of external species. However, this approach has notable limitations, such as its reliance on comprehensive knowledge of reaction mechanisms, and needs to be adapted to yield meaningful results in fluctuating environments. 

An alternative, experimentally defined notion of complexity, namely assembly theory, was introduced in \cite{sharma2023assembly}. This entails defining a set of joining operations through which molecules transform into other molecules. Then, analogously to logical depth, one can assign a complexity score to a molecule as the minimum number of joining operations necessary to construct it from the basic parts. This measure is termed the \textit{assembly index} of a molecule and can be used as the function $C$ in Eq.\ \ref{eq:complexity_abstract}. Details about the joining operations can be found in  \cite{walker2024experimentally,hazen2024molecular} and relation to other complexity measures can be found in \cite{kempes2024assembly}.

The measure of \textit{assembly} from assembly theory can also be formalized within our framework as a choice of complexity of concentrations (Eq.\ \ref{eq:complexitE_concentration}). Denoting the assembly index of a molecule $s$ as $A(s)$, the assembly of a concentration profile (Eq.\ 1 of \cite{sharma2023assembly,kempes2024assembly}) is given as
\begin{align*}
    \mathfrak{A}(q,C) &=  \sum_s  \frac{V q_s-1}{\sum_{t} Vq_{t}} e^{A(s)} \\
    & \asymp \sum_s  \frac{q_s}{\sum_{t} q_{t}} e^{A(s)},
\end{align*}
where $n_s = V q_s$ (see Eq.\ \ref{eq:conc_population}) is the copy number of a molecule $s$ in the system. Note that the assembly is an example of complexity of concentrations when the choice of the non-linear function $h(x,y)$ from Eq.\ \ref{eq:complexity_h} is 
\[h(x,y) = \sum_i \frac{x_i}{\sum_j x_j}e^{y_i}.\]

\begin{example}
    Following examples \ref{eg:1monomer_polE_model} and \ref{eg:2_1monomer_polE_model}, using Eqs.\ \ref{eq:complexity} and \ref{eq:complexitE_concentration}, we can assign measures of complexity to species and concentrations of the complete CRN $\mcl{P}'_{g,N}$. 
    
    The complexity of all external species $\mcl{E} = \mcl{E}_b$ is zero
    \[C(\mcl{E})=0.\] 
    Due to the reaction $E_0 \to \ovl{1}$, the complexity of the monomer $\ovl{1}$ is $1$
    \[C(\ovl{1}) = 1.\]
    Notice that the graph rewriting rule which corresponds to adding a new monomer to the end of the chain recovers the internal CRN.
    In particular, due to reaction $2 \ovl{1} \to \ovl{2}$, the complexity of the dimer $\ovl{2}$ is 2
    \[ C(\ovl{2}) = 2.\]
    By a similar argument, it can be seen that the complexity of $\ovl{n}$ is $n$
    \[ C(\ovl{n}) = n \hspace{1em} \forall n \in [1,N].\]
    This follows because, in this CRN, only one monomer can be added at a time. Thus the species complexity simply measures the total number of monomers that constitute the polymer (also corresponds with the conserved quantity assigned to the cluster species).

    Due to the observation above, for a concentration vector $q$ of polymers, the average complexity equals the average polymer size (see Eq.\ \ref{eq:avg_ell} and Eq.\ \ref{eq:avg_comp})
    \[ \Cpx^A(q) = \avg{\ell}_q.  \]
    Similarly, the total complexity equals the total number of monomers in the system (Eq.\ \ref{eq:total_comp})
    \[ \Cpx^T(q) = \sum_{i = 1}^N i q_i.\]

    Using $\mcl{P}_{1,4}$ and indexing the lower and higher attractors in E.g.\ \ref{eg:2_1monomer_polE_model} as $l$ and $h$, respectively, their complexity measures are
    \begin{align*}
        \Cpx^A(l) &= 1.1, & \Cpx^A(h) &= 2.4, \\
    \Cpx^T(l) &= 0.27, & \Cpx^T(h) &= 16. 
    \end{align*}
    Notice that by either measure, $\Cpx(l)< \Cpx(h)$. The complexity graph for this model is displayed in panel $b$ of Fig.\ \ref{fig:dir_evo_ex}. 
\end{example}

\subsection{Criteria for evolution}
\label{sec:evo_criterion}

\begin{figure}[t]
     \centering
  \includegraphics[width=0.93\linewidth]{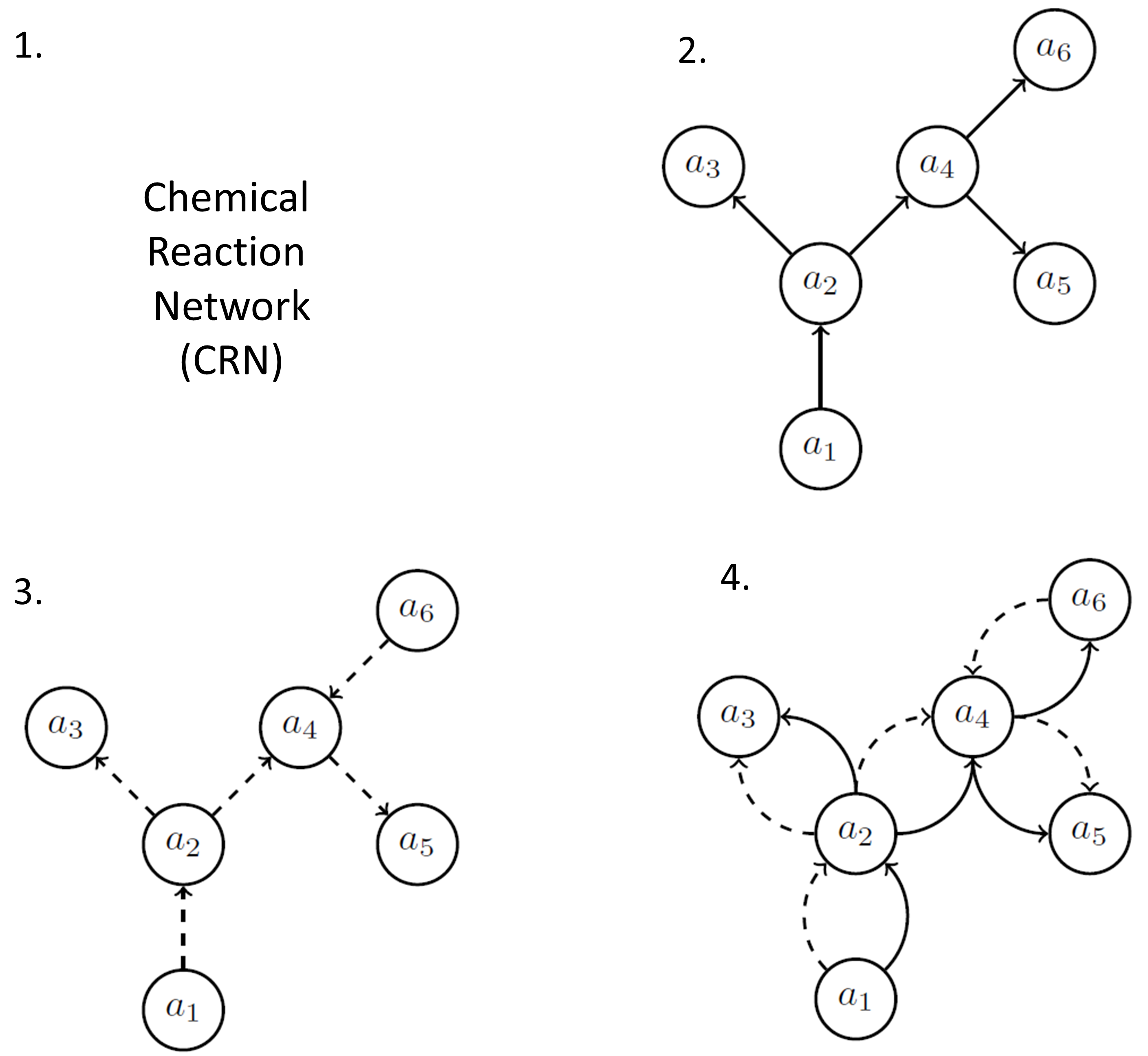}
  \caption{Steps to construct the evolution graph of a CRN. For obtaining the evolution graph for a CRN, $(1)$ start from the internal CRN, $(2)$ calculate non-equilibrium potential to extract the attractor-transition (A-T) graph, $(3)$ assign complexity to species and extract the complexity graph, $(4)$ overlay the two to get the evolution graph. In the shown graphs, the vertices are attractors (NESSs) of the system, dashed arrows represent the \textit{complexity} relationships, and the solid arrows represent the \textit{relative probabilities} of the pair of adjacent attractors under the stationary distribution. }
  \label{fig:evolution_graph}
\end{figure}

\begin{figure}[t]
     \centering
  \includegraphics[width=0.93\linewidth]{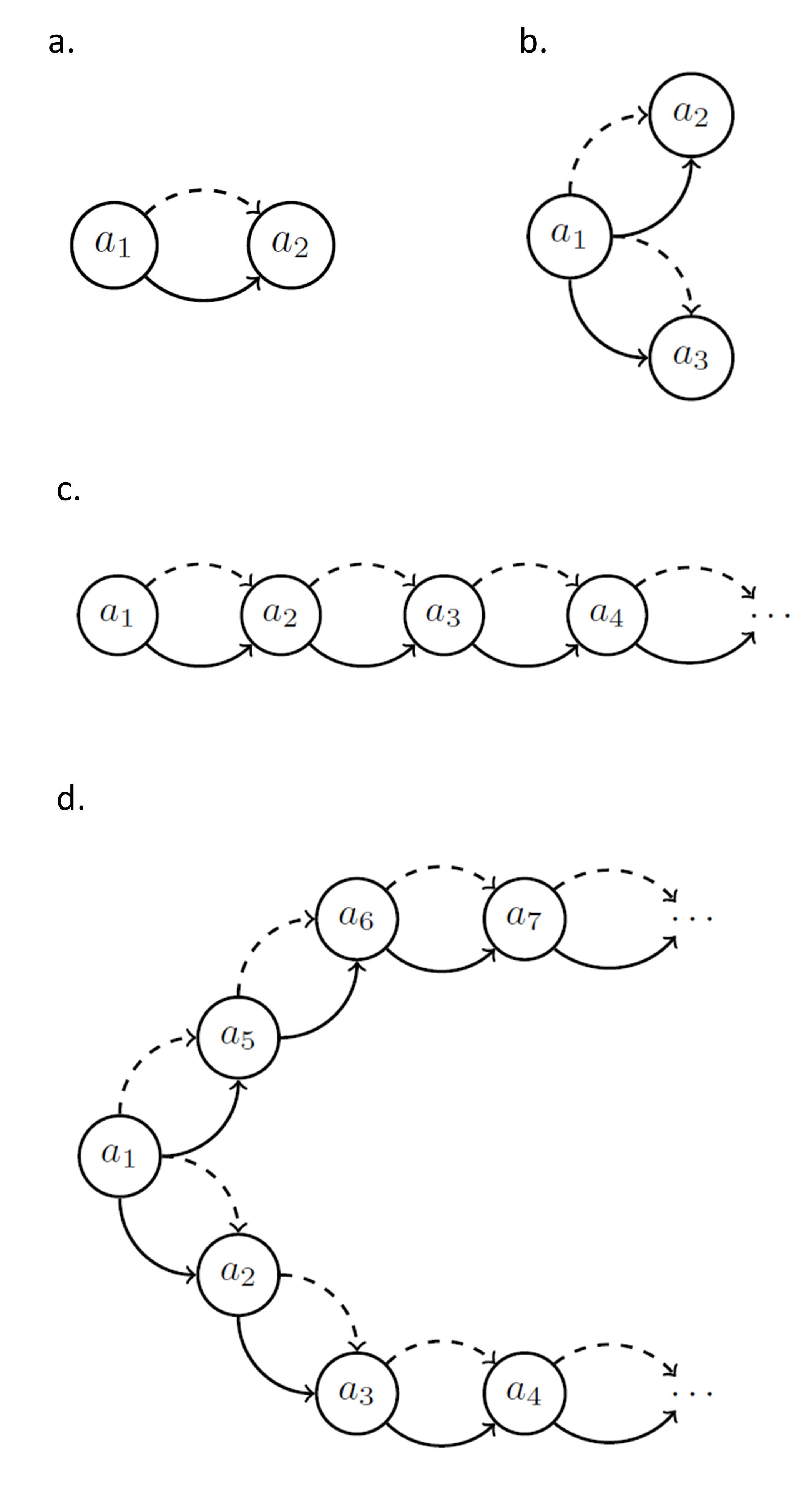}
  \caption{Visual comparison of progressive, historically contingent, and open-ended evolution criteria. Panel $a$ and $b$ display \textit{progressive evolution} and \textit{historically contingent evolution}, respectively. Panels $c$ and $d$ are both examples of \textit{open-ended evolution}, and, in particular, panel $d$ shows historically contingent open-ended evolution.}
  \label{fig:criteria}
\end{figure}

%Earlier (Sec.\ \ref{sec:internal-external}) we saw how to use a complete CRN and an external species set to obtain an internal CRN. Then (Sec.\ \ref{sec:attractor_transition}) we explained how to obtain its NESSs or attractors, estimate their probability in the stationary distribution, and encode them in a directed A-T graph. Furthermore, in Sec.\ \ref{sec:complexity}, we explained how to assign a complexity to each attractor and obtain a \textit{complexity graph}. In this subsection we explain how, using the A-T and complexity graphs, we can determine whether a CRN is capable of evolution, and, if so, of which type. 

Recall that an A-T graph is a directed graph with attractors as vertices and directed edges pointing towards an adjacent attractor with higher probability in the stationary distribution. We use \textit{solid lines} to denote the edges in the A-T graph. Furthermore, a complexity graph is also a directed graph with attractors as vertices and directed edges pointing towards a more complex adjacent attractor. We use \textit{dashed lines} to denote edges in the complexity graph. We overlay the A-T graph and complexity graph to obtain the \textbf{evolution graph} and use it to identify evolvability properties of the CRN. A graphical outline of this procedure is shown in Fig.\ \ref{fig:evolution_graph}.

In the remainder of this subsection, we specify our criteria for progressive, historically contingent, and open-ended evolution.  
\subsubsection{Progressive evolution} 
A CRN will be said to allow \textbf{progressive evolution} from an attractor if there exists an adjacent attractor that is:
    \begin{enumerate}
        \item More complex than the current attractor.
        \item More probable than the current attractor in the stationary distribution. 
    \end{enumerate} 
Graphically, as shown in panel $a$ of Fig.\ \ref{fig:criteria}, this corresponds to a pair of attractors on the graph connected by a solid and a dashed arrow in the same orientation. In that case, the CRN will be said to be capable of progressive evolution from the attractor at the start of both arrows.

\begin{example}
\label{eg:direc_evo}
    The evolution graph of the model $\mcl{P}_{1,4}$ from E.g.\ \ref{eg:2_1monomer_polE_model} is shown in the $c$ panel of Fig.\ \ref{fig:dir_evo_ex}. Since the higher attractor $a_h$ is more complex and more probable in the stationary distribution than the lower attractor $a_l$, the system is capable of progressively evolving from $a_l$ to $a_h$. Observe that the evolution graph for this model is equivalent to panel $a$ of Fig.\ \ref{fig:criteria}.
\end{example}

\subsubsection{Historically contingent evolution} A CRN will be said to be capable of \textit{historically contingent transition} from an attractor if there exist at least two adjacent attractors, such that they are:
    \begin{enumerate}
        \item More probable than the current attractor in the stationary distribution. 
        \item Not adjacent to each other. 
    \end{enumerate}
Moreover, if the two attractors are also more complex than the current attractor, then the CRN will be said to be capable of \textbf{historically contingent evolution} from the less complex attractor.

Graphically, as shown in panel $b$ of Fig.\ \ref{fig:criteria}, historically contingent evolution corresponds to having at least three attractors such that one of them is connected to the other two with a solid and a dashed edge with the tail at the first attractor, and the other two are not connected with a solid edge. 

We justify the association of the above criterion with historical contingency \cite{gould1989wonderful,blount2018contingency} because (as explained in App.\ \ref{sec:stoch_chemical_kin}) the most likely transition between attractors goes through a transient point between them. Since the more complex attractors are not directly linked by a transient point, it is highly unlikely that the system would transition directly from one of these attractors to the other without going through the less complex attractor. However, since the less complex attractor is also less probable in the stationary distribution, the transition from the more complex to a less complex attractor is also relatively unlikely. Thus, the stochastic choice made by the system in transitioning out of the attractor with lower complexity, seals the fate of the system in a probabilistic sense. This phenomenon is known as \textit{symmetry breaking} or \textit{ergodicity breaking} in physics and mathematics literature, respectively \cite{gross1996role,walters2000introduction}.

\begin{figure}[t]
\centering
  \includegraphics[width=.45\textwidth]{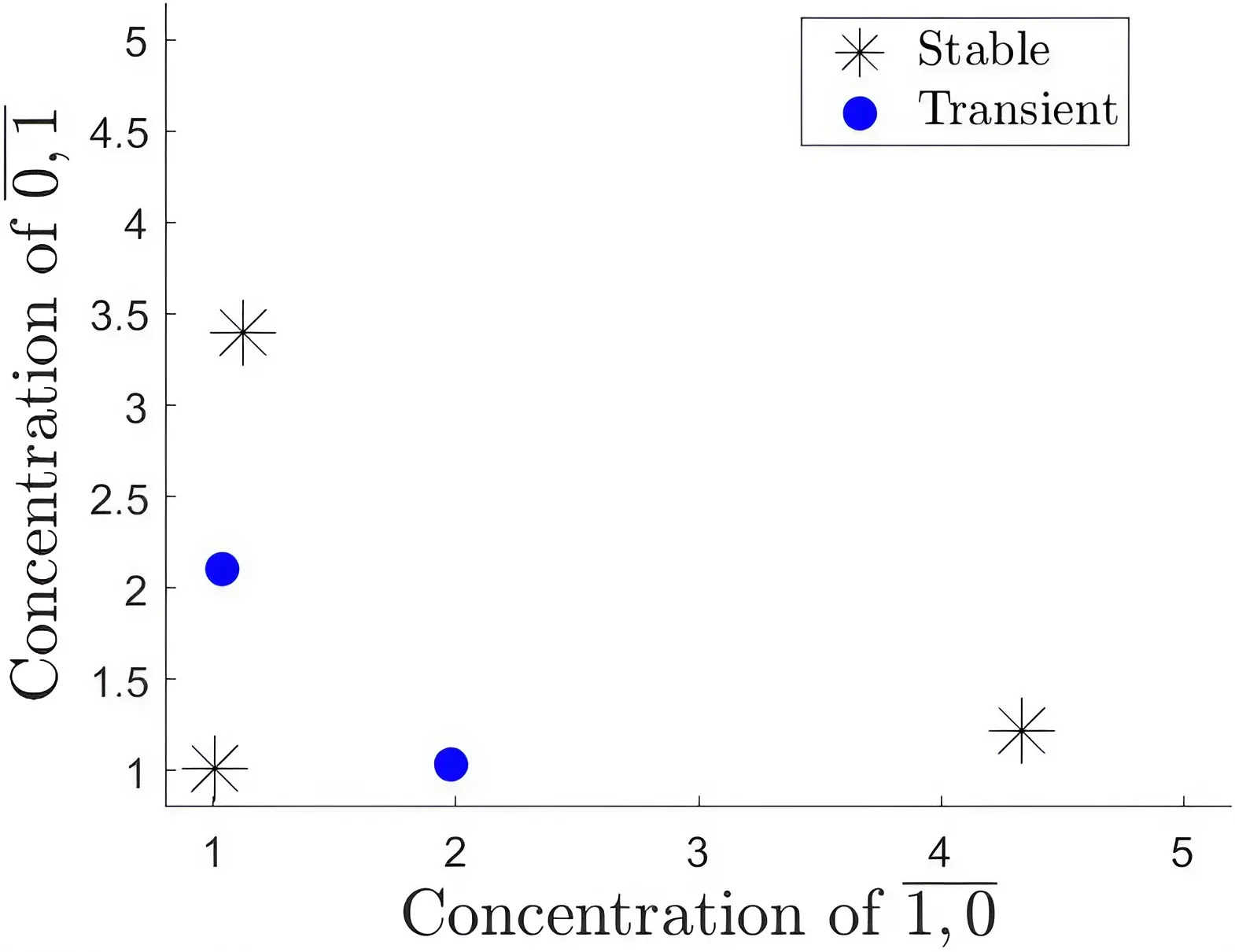}
  %\caption{Change in concentration}
\caption{ Equilibrium monomer concentrations in a 2-monomer copolymerization model, showing historically contingent evolution. As explained in E.g.\ \ref{eg:hist_cont}, there exists a rate constant assignment such that the attractors further from the origin are both more complex and more probable in the stationary distribution than the attractor closer to the origin. Furthermore, since there is no transient point between the two, the found system would be capable of exhibiting historically contingent evolution from the attractor closer to the origin.
\label{fig:hist_cont}}
\end{figure}

\begin{example}
\label{eg:hist_cont}
Consider a \textbf{2-monomer copolymerization model} \cite{andrieux2008nonequilibrium} consisting of two different types of monomers, labelled $A$ and $B$. Denoting a polymer as $\omega$, the reactions where only end additions and removals are allowed with some (polymer independent) rates $s$ and $d$, respectively, are given as:
\begin{align*}
    \ce{$\omega$ + A <=>[s][d] $\omega A$}, \hspace{1em}   \ce{$\omega$ + B <=>[s][d] $\omega B$}.
\end{align*}
Notice that the state space scales exponentially in the size of the polymer as there are $2^n$ polymers of size $n$ in two monomers. To simplify analysis, we map strings to clusters (for example, the polymer $AABA$ maps to the cluster $\ovl{3,1}$) and analyze the cluster CRN (CCRN) \cite{gagrani2024polyhedral} associated with the copolymerization model, as described below.

The species set of a two constituent CCRN is given by
\[\mathcal{S} = \{ \overline{1,0},\overline{0,1},\overline{1,1},\ldots,\overline{i,j},\ldots\},\]
for $i,j \geq 0 \in \mathbb{Z}$. The internal-internal reaction set induced by end additions and removals is given by
\begin{align*}
  \mathcal{R}_i =&  \bigg\{ \ce{\overline{i,j} + \overline{1,0} <=>[s][d] \overline{i+1,j}}\hspace{0.1em}, \hspace{0.2em} \ce{\overline{i,j} + \overline{0,1} <=>[s][d] \overline{i,j+1}}  \bigg|\\
  & \phantom{=} \forall i,j \in \mathbb{Z}_{\geq 0},  N \geq i+j\geq 1\bigg\},
\end{align*}
where $s$, $d$, and $N$ are rates of addition, removal, and maximum polymer length, respectively. 

Analogous to the internal CRN of the 1-monomer polymerization models from E.g.\ \ref{eg:1monomer_polE_model}, we introduce reactions for monomer production and catalyzed monomer production. We assume that the clusters $\ovl{2,0}$ and $\ovl{3,2}$ catalyze the production of $\ovl{1,0}$, and $\ovl{0,2}$ and $\ovl{2,3}$ catalyze the production of $\ovl{0,1}$. A concrete realization of the model with rate constants is given by the reaction set: 
\begin{align*}
    \mathcal{R} = & \mathcal{R}_i|_{s=1, d= 5.01} \cup \\
    & \bigg\{ \hspace{0.5em} \ce{ $\overline{0,1}$ <=>[17][10] $\varnothing$ <=>[10][17] $\overline{1,0}$} \hspace{0.3em}, \\
    & \phantom{=} \ce{$\overline{2,0}$ <=>[40.08][5.01] $\overline{2,0} + \overline{1,0}$ } \hspace{0.3em}, \hspace{0.5em} 
    \ce{$\overline{3,2}$ <=>[31.5][13.54] $\overline{3,2} + \overline{1,0}$ },\\
    & \phantom{=} \ce{$\overline{0,2}$ <=>[40.08][5.01] $\overline{0,2} + \overline{0,1}$ } \hspace{0.3em}, \hspace{0.5em}
    \ce{$\overline{2,3}$ <=>[56.7][36.82] $\overline{2,3} + \overline{0,1}$ }\bigg\}.
\end{align*}
The attractors in the monomer concentrations are shown in Fig.\ \ref{fig:hist_cont}. It can be seen that the CRN has three attractors, roughly around $(1,1),(4,1)$ and $(1,3)$, and we denote them as $a_1$, $a_2$, and $a_3$, respectively. 

Notice that while there is a transient point between $a_1$ and $a_2$, and $a_1$ and $a_3$, there is no transient point between $a_2$ and $a_3$. This means that $a_1$ is adjacent to both $a_2$ and $a_3$, however $a_2$ and $a_3$ are not adjacent to each other. By making the transient points much closer to $a_1$ than $a_2$ or $a_3$, as explained in Sec.\ \ref{sec:attractor_transition} and illustrated in E.g.\ \ref{eg:2_1monomer_polE_model} and \ref{eg:ex4_3models_gen_schlogl}, the NEP at $a_1$ can be made higher than both $a_2$ or $a_3$. Thus, there exists an assignment of rate constants such that $a_2$ and $a_3$ are more probable than $a_1$ in the stationary distribution. Furthermore, (by the same arguments as in E.g.\ \ref{eg:2_1monomer_polE_model}) the polymer concentration in the equilibrium distribution follows an exponential distribution and the average polymer length as well as the total monomer concentration in $a_2$ and $a_3$ is higher than in $a_1$. Thus, using our complexity measure for concentrations (Sec.\ \ref{sec:complexity}), $a_1$ is the least complex attractor.
This means that there exists a rate constant assignment such that $a_2$ and $a_3$ are more complex and probable in the stationary distribution than $a_1$, and also not adjacent to each other. The evolution graph for such a system is equivalent to panel $b$ in Fig.\ \ref{fig:criteria} and the system can exhibit historically-contingent evolution from $a_1$.
\end{example}

\subsubsection{Open-ended evolution} A CRN will be said to be capable of \textit{open-ended evolution} if there is at least one subset of infinitely many attractors such that each one is capable of progressive evolution to another attractor in the subset. 

Some graphs displaying open-ended evolution are shown in panels $c$ and $d$ of Fig.\ \ref{fig:criteria}. Notice that there is only one open-ended direction for the system shown in panel $c$, however panel $d$ has two historically-contingent directions for exhibiting open-endedness.

The number of possible molecular configurations constitutes a vast combinatorial space, making the enumeration of all attractors infeasible and largely uninformative. The idea that an evolving system’s present state strongly constrains its future possibilities is well established in the literature—see, for example, the concept of the adjacent possible \cite{kauffman2014prolegomenon} and the role of self-referential dynamics in evolution \cite{goldenfeld2011life}. In the remainder of this subsection, we argue that the proposed framework is sufficiently rich to capture dynamics relevant to extant life. 

First, stochastic CRNs can exhibit a wide range of dynamic behaviors, including limit cycles, chaotic attractors \cite{yu2018mathematical}, and transient states \cite{norris1998markov}. In particular, transience in CRNs has been used to model indefinitely growing polymer systems \cite{andrieux2008nonequilibrium,gaspard2016kinetics,esposito2010extracting}. While our current analysis focuses on point attractors, future work could extend these conclusions to other types of attractors and transient behaviors.

Second, to analyze the dynamics of complex scenarios within a combinatorially vast space, our framework proposes an iterative approach: beginning from an attractor, constructing its evolution graph numerically, transitioning to adjacent attractors, and repeating the process. We posit that a system's ability to exhibit open-ended evolution requires that the above algorithm never halts. When the explored space is a rule-generated CRN, the methods outlined in this paper provide a precise framework for conducting such an analysis.

\begin{example}
\label{eg:open_ended}
    As explained earlier (in E.g.\ \ref{eg:2_1monomer_polE_model}), $\mcl{P}_{g,N}$, with $N \geq 2g$, can have up to $2g + 1$ fixed points and $g+1$ attractors. Furthermore, by the arguments in Sec.\ \ref{sec:attractor_transition}, a rate constant assignment can always be made such that the NEP of the more complex adjacent attractor is also lower, thus ensuring that the more complex attractor is more probable in the stationary distribution. A one dimensional analog of the model with $3$ attractors is shown in E.g.\ \ref{eg:ex4_3models_gen_schlogl}. 
    
    Note that the extension of this model to infinite positive attractors is technically challenging as the rate constants also increase, raising issues of summability and convergence. Thus, while we propose a criterion for open-ended evolution, a proper treatment of reaction networks with infinite reactions and fixed-points will be left as future work.
\end{example}

\begin{example}
Applying the criteria explained in this subsection to the evolution graph displayed in Fig.\  \ref{fig:evolution_graph} panel 4 panel $(4)$, it can be seen that the system is: 
\begin{itemize}
    \item Capable of progressive evolution from attractor $a_1$, $a_2$, $a_4$, $a_5$.
    \item Capable of historically contingent evolution from attractor $a_2$.
    \item Incapable of open-ended evolution because of a finite number of attractors.
\end{itemize}
\end{example}

%\end{comment}    

\section{Remarks on framework and model}
\label{sec:remarks}

%\begin{comment}
    
In this section, we remark on the roles of entropy production rate (EPR) and autocatalysis in our framework and models. In Sec.\ \ref{sec:remarks_EPR_evolution}, we explain the indirect role of EPR in our definition of evolution. In Sec.\ \ref{sec:math_form}, we defined a class of polymerization models and demonstrated that a model with $g$ polymer catalysts can exhibit up to $g+1$ attractors. For these models, in Sec.\ \ref{sec:autocat_remarks}, we explain the role of autocatalysis in making multistability possible. In Sec.\ \ref{sec:relationship_pop_gen}, we outline how terminology in our framework maps onto key concepts in evolutionary biology and consider some implications for the origins of life.

\subsection{Role of dissipation and efficiency in framework}
%\end{comment}

\label{sec:remarks_EPR_evolution}

\begin{comment}    
\begin{figure}
    \centering
    \includegraphics[width=.45\textwidth]{figures/EPR_avgpoly.png}
    \caption{Caption}
    \label{fig:EPR_polE_model}
\end{figure}
\end{comment}

%\begin{comment}

In \cite{england2015dissipative}, using fluctuation theorems concerning dissipation and irreversibility proposed in \cite{jarzynski1997nonequilibrium,crooks1999entropy}, England suggests a relationship between biological organization and EPR (for fluctuation theorems in CRNs, see \cite{schmiedl2007stochastic}). In particular, it is conjectured that biological systems evolve to maximize entropy production or dissipation. Similarly, in \cite{kondepudi2020dissipative}, complex systems are found where the system temporally evolves to states of higher EPR. In this subsection, we briefly summarize the role played by EPR and free-energy transduction efficiency in our framework. (See App.\ \ref{sec:thermo_CRN} for a review of related concepts.)

First, one can show that the EPR only informs about the stationary probability distribution of the system when the system admits a single detailed-balanced or complex-balanced equilibrium in each stoichiometric compatibility class (see discussion after Eq.\ \ref{eq:EP}). In particular, the expression for EPR and the integrand of the NEP with time as the variable of integration coincide (up to sign) only for detailed-balanced systems. Since we are primarily interested in multistable systems with many NESSs, the EPR cannot be used to determine relative probabilities of the NESSs or transitions between them. Thus, EPR cannot be used to obtain an attractor-transition graph (see Sec.\ \ref{sec:attractor_transition}).

Second, in Sec.\ \ref{sec:complexity}, we explained how internal reactions can be obtained through the application of graph-rewriting operations, and used them to assign a complexity measure to species that induces a complexity measure on the NESSs. However, because EPR at a NESS only depends on the chemical potential and fluxes of the external boundary species (see Eq.\ \ref{eq:EPR_emergent}), it does not depend on the internal-internal species that make up the equilibrium concentration profile of the NESS. Hence, EPR cannot be used to obtain a complexity measure on the NESSs either. 

In summary, the EPR at NESSs need not always be correlated with either their probability in the stationary distribution or their complexity. Nonetheless, as reviewed in App.\ \ref{sec:thermo_CRN}, one can decompose a CRN into processes and obtain a free-energy transduction efficiency using the EPRs of reactions at each of the NESSs. Thus, in principle, the efficiency (or EPR) could be used as a measure for ranking the NESSs and obtaining a dissipation analog of a complexity graph. The relevance of such a ranking to the origins of life problem is, however, unclear.

\begin{example}
   As shown in E.g.\ \ref{eg:2_1monomer_polE_model} and \ref{eg:ex4_3models_gen_schlogl}, the steady state behavior of the 1-monomer polymerization models from E.g.\ \ref{eg:1monomer_polE_model} is identical to the complete generalized Schl\"{o}gl model introduced in E.g.\ \ref{eg:emb_gen_schlogl_defnt}. Since the efficiency at NESSs only depends on the steady state behavior, the efficiency analysis of the polymer model is identical to that of the Schl\"{o}gl model. 
   
   An efficiency analysis of NESSs of the generalized Schl\"{o}gl model, shown in E.g.\ \ref{eg:schlogl_efficiency}, demonstrates that (Fig.\ \ref{fig:transduction_efficiency}) the efficiency increases monotonically in the distance of the attractor from the origin and does not correlate with the NEP or the probability of the attractor in the stationary distribution. Although this example is a case where the efficiency at the various attractors is positively correlated with their complexity, unless a theorem is proved otherwise, this cannot be assumed to always be the case.

\end{example}

\subsection{Role of autocatalysis in models}
\label{sec:autocat_remarks}
In Sec.\ \ref{sec:math_form}, we defined a class of polymerization models and demonstrated that a model with $g$ polymer catalysts can exhibit up to $g$ attractors. 
In this section, we explain that these polymer catalysts are \textit{autocatalysts}, and a model with $g$ polymer catalysts has exactly $g$ minimal autocatalytic subnetworks. We term the species like the catalysts as \textit{functional species} and elaborate on conditions satisfied by the models that yield multiple attractors.

\subsubsection{Review: Autocatalysis}
\label{sec:autocatalysis_review}
Autocatalysis is an important property of CRNs that generalizes the biologically relevant notions of self-replicability and ecology \cite{deshpande2014autocatalysis,blokhuis2020universal,peng2020ecological,gagrani2024polyhedral}. In \cite{peng2020ecological}, it is shown that different autocatalytic subnetworks can be composed to achieve a variety of ecologically relevant dynamical behaviors including competition, predator-prey, and mutualism. In \cite{munuzuri2022unified,plum2022aces}, it is shown that autocatalytic subnetworks on a spatial lattice exhibit Turing patterns and other spatial patterns relevant to the origins of life. The kinetic role of autocatalysis is explained in \cite{vassena2024unstable}, and some thermodynamic results about autocatalysis are shown in \cite{despons2023structural}. 

Let $\mcl{G} = (\Spc,\R)$ denote a subnetwork of $\mcl{G}'= (\Spc',\R')$, where $\R \subset \R'$ and $\Spc$ consists of all the species of $\Spc'$ that participate in $\R$. For $\mcl{G} = (\Spc,\R)$, we denote the input, output, and stoichiometric matrix as $\St^-$, $\St^+$, and $\St$, respectively (see App.\ \ref{sec:CRN_intro}), and the restriction of a $\Spc \times \R$ matrix $\mathbb{M}$ to a set $\mcl{M} \subset \Spc$ as $\mathbb{M}|_{\mcl{M}}$. The reactions in the complement of $\mcl{G}$ are given by all reactions in $\R'$ that are not in $\R$ and denoted as $\R'/\R$.

A subnetwork is autocatalytic in the \textit{autocatalytic set} if there exists a reaction flow such that every species in the autocatalytic set is consumed but also net-produced.
Formally, a subnetwork $\mcl{G} = (\Spc,\R)$ is \textbf{autocatalytic} in the species set $\mcl{M}\subseteq \Spc$ if (see \cite{gagrani2024polyhedral} and references therein):
\begin{itemize}
    \item \textbf{Autonomy:} Every species in $\mcl{M}$ is consumed and produced by at least one reaction in $\R$, and every reaction consumes and produces at least one species in $\mcl{M}$,
    \begin{align*}
        \forall s \in \mcl{M}, \exists r,r' \in \R : &\quad (\St^{-})^r_s > 0 \text{ and } (\St^{+})^{r'}_s > 0,\\
        \forall r \in \R, \exists s,s' \in \mcl{M} : &\quad (\St^{-})^r_s > 0 \text{ and } (\St^{+})^r_{s'} > 0.
    \end{align*}
    \item \textbf{Productivity:} There exists a semi-positive reaction flow vector such that all species in $\mcl{M}$ are net produced, (see Sec.\ \ref{sec:basic_terminology} for notation)
    \[ \exists \mbf{x} \in \mathbb{R}^{\R}_{\geq 0} : \quad \St|_{\mcl{M}} \mbf{x} \gg \mbf{0}.\]
\end{itemize}
Under these conditions, $\mcl{M}$ is an \textbf{autocatalytic set}. The restriction of the stoichiometric matrix to the autocatalytic set  $\mathbb{S}\big|_\mathcal{M}$ is a semi-positive (SP) matrix (Ch.\ 3, \cite{johnson_smith_tsatsomeros_2020}).
If each $\mbf{x}$ that maps in the productive region is strictly positive, or $\kappa \gg \mathbf{0}$, then $\mathbb{S}\big|_\mathcal{M}$ is minimally SP (MSP) and the reaction network $\mathcal{G}$ is minimally productive in $\mathcal{M}$. By definition, no reaction-reduced subnetwork of a minimal productive network can be productive in the autocatalytic set. One can further restrict the set of autonomous species to find a minimal autocatalytic motif, also called an autocatalytic \textbf{core}. It can be shown that the stoichiometric matrix of a (autocatalytic) core is an invertible matrix \cite{blokhuis2020universal}. We refer to the species in the core as \textbf{core species}. 

\subsubsection{Heuristics for multistability}

There are two aspects of multistability: kinetic and topological.

With regards to the kinetic requirement for multistability, recall from Sec.\ \ref{sec:internal-external} that for thermodynamic consistency, the rate constants on the complete network are assumed to be such that the system has a single detailed-balanced equilibrium in each stoichiometric compatibility class. However, as explained below Eq.\ \ref{eq:rate_const_renorm}, by introducing an environmental control in the form of chemostatted external-boundary species, the rate constants of the system can depart from a detailed-balanced equilibrium. Thus, the first heuristic for multistability is that there must be at least one boundary reaction in the complete CRN.

In regards to the topological requirement for multistability, non-zero deficiency (Eq.\ \ref{eq:deficiency}) is a necessary condition for multistability in a reversible CRN (see App.\ \ref{sec:CRN_intro}). The appendix also explains that the deficiency, $\delta$, of a CRN counts the number of null flows, termed $\mbf{\delta}$\textbf{-flows}, that are not supported in a single linkage class. From Sec.\ \ref{sec:autocatalysis_review}, recall that each autocatalytic subnetwork has a core whose stoichiometric matrix is invertible. This means that there exists a reaction-flow vector through the autocatalytic subnetwork that creates exactly one core species. If there exists a reaction-flow in the complement of the core that also produces any of the core species, and if the above mentioned two reaction-flows have a support in at least two linkage classes, then the deficiency of the network must be non-zero. Thus, the presence of autocatalytic cores is our second heuristic for multistability. 

To summarize, the existence of at least one boundary reaction and at least one autocatalytic core are our two heuristics for multistability. We term an internal-boundary species that is also a core species a \textbf{functional species}. Such a species can perform the dual role of transducing free-energy from the environment and super-linear growth in some concentration regimes, satisfying a few conditions necessary for multistability.

\begin{example}
    Consider a subnetwork $\mcl{H} = (\Spc_\mcl{H},\R_\mcl{H})$ of the internal CRN $\mcl{P}_{1,2}$ from E.g.\ \ref{eg:1monomer_polE_model}
    \begin{align*}
        \Spc_\mcl{H} &= \{ \ovl{1},\ovl{2}\},\\
        \R_\mcl{H} &= \{ 2\ovl{1} \to \ovl{2}, \ovl{2} \to \ovl{1}+\ovl{2}\}.
    \end{align*}
    Observe that $\mcl{H}$ is autonomous in $\Spc_\mcl{H}$, as both species are produced and consumed and each reaction produces and consumes both species. Moreover, the network is also productive as all the species are net produced under the flow $\mbf{x} = [1,3]^T$,
    $$\St \mbf{x} = \begin{bmatrix}
        1\\
        1
    \end{bmatrix} \gg \mbf{0}.$$ Thus, $\mcl{H}$ is an autocatalytic subnetwork. Furthermore, due to the reaction $\ovl{1} \to \vn$ in the CRN $\mcl{P}_{1,2}$, the autocatalytic subnetwork produces a $\delta$-flow
    $$ \ovl{2} \to \ovl{1} + \ovl{2}, \quad \ovl{1} \to \vn.$$

    As explained in E.g.\ \ref{eg:2_1monomer_polE_model}, due to the external species in the network, $\ovl{1}$ and $\ovl{2}$ are both internal-boundary species. Since they are both internal-boundary species and autocatalytic species, we call them \textit{functional species}. We have shown that there exist rate constants such that $\mcl{P}_{1,2}$ exhibits multistability, and we can attribute it to the presence of functional species. By the same argument, one can see that for a model $\mcl{P}_{g,N}$, there are $g$ independent autocatalytic subnetworks and $g+1$ functional species. Moreover, each autocatalytic subnetwork creates an independent $\delta$-flow, and there exist rate constants such that the system exhibits $g+1$ attractors.    
\end{example}

%\end{comment}

\subsection{Relationship to population genetics}
\label{sec:relationship_pop_gen}

In conventional evolutionary biology, the concept of evolution is generally defined as a change in the frequency of a gene variant, an allele, in a population of organisms. In this framing, allele frequency change is considered evolution regardless of the cause of this change, specifically whether it happens by chance (genetic drift) or by selection, where alleles differ in their effects on the expected reproductive output (fitness) of organisms. In this subsection, we explain how our framework can be applied to the origins of biochemical life and maps onto key concepts in evolutionary biology. 

\begin{figure}
    \centering
    \includegraphics[width=.5\textwidth]{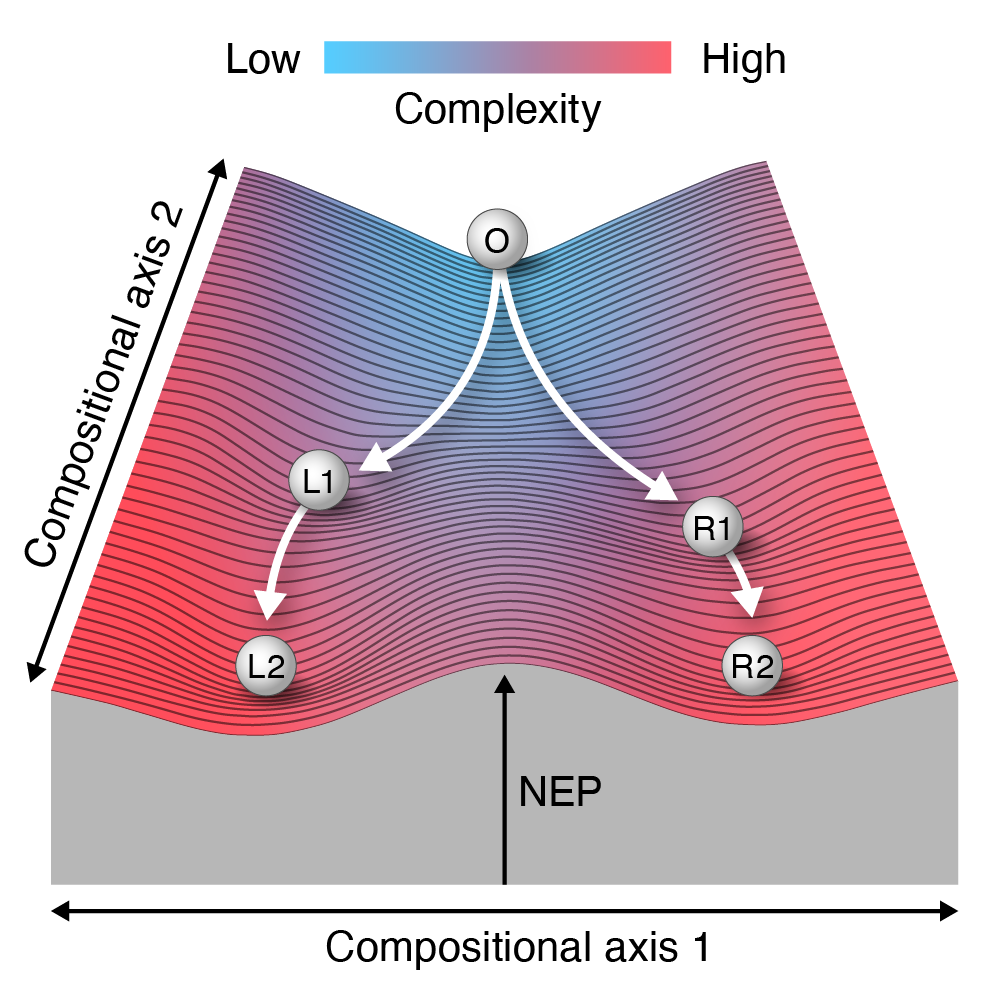}
    \caption{Simplified attractor landscape for a 2-monomer polymerization model and its connection to population genetics. The details of the model and its relevance to population genetics is explained in Sec.\ \ref{sec:relationship_pop_gen}.}
    \label{fig:waddington}
\end{figure}

Consider a 2-monomer polymerization CRN (with catalysis, similar to Ex.\ \ref{eg:hist_cont}) whose landscape is shown qualitatively in the Fig. \ref{fig:waddington}. While the population is very high dimensional, we show a simplified representation on a polar coordinate chart. The angular coordinate, compositional axis 1, measures the ratio of the abundance of the two monomers in the population, whereas the radial coordinate, compositional axis 2, is the average polymer length of the population. The CRN can exhibit 5 attractors $O, L1, L2, R1, R2$ where the NEP of $L2<L1<O$ and $R2<R1<O$, and any transition from $L1$ to $R1$ is exponentially likely to happen through $O$. Since the complexity of a population is aligned with compositional axis 2 (depicted by a color gradient from blue to red), the population with a higher average polymer length will be said to be more complex than another with a lower average population. In our terminology, this system is capable of historically contingent evolution from attractor $O$ and progressive evolution from attractors $L1$ and $R1$ to $L2$ and $R2$, respectively.

To help draw a connection to modern cells, let us assume that the CRN is instantiated in a solution that can exhibit spontaneous phase transitions leading to the formation of particles in the solution, which we will call protocells. Let us further suppose that protocells can grow and divide spontaneously as a function of their polymer biomass. The mathematics of compartmentalized chemical systems has been formalized in \cite{duso2020stochastic,anderson2023stochastic} and its implications for the origins of life have been considered in the GARD model \cite{segre1998graded}. Thermodynamics of CRNs in growing systems has also been studied \cite{sughiyama2022chemical}, and it is known that division cycles can alter the attractor landscape \cite{matsubara2022robustness}. If we assume that the time scale of relaxation to an attractor is much shorter than the time scale of division, each protocell can be assigned a state $O$, $L1$, $L2$, $R1$ or $R2$ corresponding to the attractors, as labeled in the figure.

Each time a protocell splits into two, the concentration of the offspring protocells has the potential to be initialized in a different basin of attraction than the parental protocell. As explained in Sec.\ \ref{sec:attractor_transition}, in general, to make the more complex attractor more probable, the saddle point should be closer to the less complex attractor. Thus, we expect that at least one of the offspring protocells will be in the same attractor as the parent. For example, if a protocell $L1$ splits in two, at least one of them is very likely to be $L1$. This property of the model maps onto \textbf{heritability}. Notice that there is a possibility that one daughter cell is initialized closer to another attractor meaning it might transition to that adjacent attractor. For example, a parental protocell of type $L1$ could reproduce into two protocells of types $L1$ and $O$. Additionally, transitions could occur stochastically during a protocell’s growth. In both cases, such transitions may be denoted as \textbf{mutations}. 
%must entail changes in flux through different reaction channels. The reaction flux changes responsible for a transition may be denoted mutation reactions. 

Two factors can impose directionally on system behavior. First, we should expect mutations to be more frequent from high to low NEP attractors than the reverse, such that a protocell population should be driven down contingent channels towards lower NEP states. Second, we should expect populations to be dominated by states that are associated with higher rates of growth and division. Supported by the observation that the entropy production rate at more complex attractors are higher (see App.\ \ref{sec:thermo_CRN}), we should expect the growth rate at more complex attractors to also be higher as they intake more monomers from the environment. Thus, we might expect \textbf{selection} to favor more complex protocells. 
%When the first factor, mutation bias, tends to favor different attractors than the second factor, selection, for a sufficiently complex attractor, we can expect selection to dominate. This is because mutation rates will be much lower than the rate of protocell division as they require rare escape trajectories from one basin of attraction into another.

%[If you can prove that more complex attractors should have higher reproductive rates = protocell fitness = this would be the place to insert that. If you do, that would be a big deal and imply a try ratchet towards complexity!]

The preceding description closely resembles the GARD model of compositional heredity and adaptive evolution \cite{lancet2018systems,kahana2023attractor}.  However, by referring to a CRN, specifically a polymer CRN, it is easier to see the connections to genetic evolution. Such a CRN will contain, generically, many overlapping autocatalytic subnetworks sharing the same species, with different autocatalytic channels dominating in the vicinity of different attractors. For instance, autocatalytic networks involving more $L$ and $R$ monomers will be dominant near the attractor $L2$ and $R2$, respectively. Polymers with a high number of $L$’s will be present at a high concentration in $L2$, while their population will be exponentially suppressed in the attractor $R2$. What this means is that when a mutation occurs, one or more autocatalytic subnetwork increases its flux and one or more decrease their flux. Moreover, as a system complexifies, there may be certain autocatalytic species that are dominant in one attractor while completely absent in another. Such a situation converges towards the case of a genetic cell, where each attractor is dominated by a polymer of a particular sequence, an \textbf{allele}, with mutations being rare reactions that interconvert one allele into another. Each allele is a particular chemical species that is a member of a different autocatalytic subnetwork within cellular metabolism – specifically an autocatalytic motif that uses template-guided replication to make additional copies of molecules of that allele. It is worth emphasizing that heritability, the capacity of alleles to be passed on to offspring, is a direct manifestation of this autocatalysis, just like in our model.

Genetic cells correspond to autocatalytic chemical ecologies that include multiple cooperating, complex polymers, enabling open-ended evolution. While these may initially appear qualitatively distinct from simple autocatalytic CRNs, our description outlines a continuous path linking these two seemingly distinct kinds of chemical ecology. It seems to us that understanding this transition demands that we explain how multistable CRNs come to enter regions of state space in which the domains of attraction are labeled by binary counts of individual chemical types (alleles), where these molecules are themselves linked by exponentially many distinct mutational channels. While there is much more work to do, by rigorously defining key concepts such as ecology, evolution, and complexity in the context of stochastic CRNs, this paper can be seen as an important step towards explaining how genetically evolving life came to be.

%\begin{comment}
    
\section{Discussion}
\label{sec:discussion}

In this work we have formalized phases in CRNs under mass-action kinetics (MAK) as \textit{attractors} (stable fixed points). Although closed and thermodynamically consistent systems can only exhibit a single detailed-balanced attractor, open systems can exhibit multistability (i.e., multiple attractors or non-equilibrium steady states). We propose that multistability is a precondition for evolution and define four steps to ascertain whether a multistable CRN is capable of progressive evolution. First, start with a complete CRN and obtain its internal-boundary-external partition by specifying the species that are internal to the system under investigation and its environment. We restrict this work to scenarios where the environmental species are chemostatted and refer to the resulting network as the internal CRN (Sec.\ \ref{sec:internal-external}). Second, determine the attractors for the internal CRN and their adjacencies using MAK, and assign to each attractor a stability using its non-equilibrium potential (NEP) (App.\ \ref{sec:stoch_chemical_kin}). This information can be graphically represented in an \textit{attractor-transition graph}  (A-T graph) with vertices as attractors, and directed edges between adjacent attractors pointing towards the more stable attractor (Sec.\ \ref{sec:attractor_transition}). Third, superimpose a \textit{complexity graph} on the A-T graph with the same vertices and edges, but where the edge direction is from less complex to more complex adjacent attractors. This step requires a (partial) ordering of the attractors by their complexity, which we define using measures from algorithmic complexity (Sec.\ \ref{sec:complexity}). Finally, we superimpose the two graphs to obtain an \textit{evolution graph} and specify that a CRN is capable of progressive evolution from an attractor if there exists another attractor adjacent to it that is more stable and more complex (Sec.\ \ref{sec:evo_criterion}).

In Sec.\ \ref{sec:remarks_EPR_evolution}, we explain that evolution does not always correspond with an increase in entropy production rate or increase in free-energy transduction efficiency of the attractors. A further technical contribution of this work lies in defining the concept of a \textit{functional species} and exploring its connection to multistability in Sec.\ \ref{sec:autocat_remarks}. A functional species has two key roles, namely: (1) interacting with the environment to transduce chemical potential energy into the system needed to maintain it at a non-detailed balanced attractor, and (2) autocatalysis in the internal species. Within a range of rate constants for our models, these interactions introduce additional attractors at higher average polymer lengths, while also enforcing that the attractor with a higher average length is also more stable. By a similar construction, in Sec.\ \ref{sec:evo_criterion}, we show that adding a functional polymer species in a two monomer polymerization CRN can make the system capable of historical contingency, and that adding a countable number of functional species can, in principle, make the system open-ended. In Sec.\ \ref{sec:relationship_pop_gen}, we describe how a CRN, when implemented in a solution exhibiting spontaneous phase separation, produces a system that parallels characteristics of existing life such as heredity, mutation, and selection. Thus, we demonstrate that, at least mathematically, one can find physically realizable CRN models that exhibit behaviors that resemble aspects of biochemical life.  

Our work has several direct implications for the study of the origins of biochemical life. Given a prebiotically plausible CRN with rate constants, the proposed framework can be used to numerically ascertain whether the CRN is capable of progressive, historically-contingent, or open-ended evolution under given environmental conditions. The distribution of attractors and their relative stabilities can be found using numerical solvers for finding attractors (such as Bertini \cite{BHSW06}) and algorithms for estimating the NEP between them \cite{gagrani2023action}. The concept of an evolutionary graph could readily be employed beyond mass-action to other kinetic models such as Michaelis-Menten \cite{qian2002single}, used to account for modes of environmental control from the environment other than chemostatting \cite{kim2020absolutely}, for protocols that deviate from continuously-stirred tank reactors (such as serial-dilution experiments \cite{matsubara2022robustness}), or to incorporate other types of dynamics such as limit cycles, chaotic attractors \cite{strogatz2018nonlinear}, and transience \cite{norris1998markov}. 

By combining advanced techniques in probability theory with foundational concepts in evolutionary biology, we have rigorously defined the concept of evolution for stochastic CRNs. Similar to the Gibb's free energy in statistical mechanics \cite{pathria2016statistical}, our framework relies on the NEP for stochastic processes \cite{anderson2015lyapunov} to identify attractors and their relative stabilities. Our approach suggests that if all biologically relevant phenomena can be formalized in a stochastic framework, then there is a natural path of complexification from prebiotic chemistry to modern life \cite{baum2023ecology}, with foundations in non-equilibrium statistical physics.

\section*{Acknowledgements}

 This work was funded by the National Science Foundation, Division of Environmental Biology (Grant No: DEB-2218817). We thank  Christoph Flamm, David Anderson, Diego Rojas La Luz, Massimo Bilancioni, Massimiliano Esposito, and Atsushi Kamimura for technical conversations and references, Vladimir Sotirov and Eric Smith for help in conceptualizing and clarifying several aspects of this work, Sarah Friedrich for assistance with Figures 3 and 8, and Helen Noeldner for recommending copy edits.    

\appendix

\section{Mathematical preliminaries}
\label{sec:math_prelim}

\subsection{Basic terminology}
\label{sec:basic_terminology}
\subparagraph{Scalars}
\begin{itemize}
\item The set of real numbers and its subset of integers are denoted by
	$\mathbb R$ and $\mathbb Z$.
\item For any subset $X\subseteq\mathbb R$ of real numbers, its subsets of
	non-negative and positive elements are denoted by
	$X_{\ge 0}=\{x\in X:x\ge 0\}$ and $X_{>0}=\{x\in X:x>0\}$.
\end{itemize}

\subparagraph{Vectors}
\begin{itemize}
\item $\mbf{0}$ denotes a vector of zeros.
\item Given any sets $X$ and $Y$, the set of $Y$-valued functions with domain
	$X$ is denoted by $Y^X$.

\item A real-valued function with domain a set $X$, i.e., an element
	$v\in\mathbb R^X$, is called an $X$-\textbf{vector}.

\item The value at $x\in X$ of $\mbf{v}\in\mathbb R^X$ is its 
	$x$-\textbf{component} and is denoted by $\mbf{v}_x$.

%\item The \define{support} of an $X$ vector $ v\in\mathbb R^X$ is
%	the set $\supp v=\{x\in X: v_x\ne 0\}$.

 \item Given two $X$ vectors $a$ and $b$, we use $a^b$ to denote $\prod_{i\in X} a_i^{b_i}$.
\end{itemize}

\subparagraph{Matrices}
\begin{itemize}
\item A $\mathcal S\times\mathcal R$ \textbf{matrix} is a two-variable
	function with domain a pair of finite sets $(\mathcal S ,\mathcal R)$,
	i.e., an element $\mathbb{M}\in\mathbb R^{\mathcal S\times\mathcal R}$.

\item The value at $(s,r)\in\mathcal S\times\mathcal R$ of 
	$\mathbb{M}\in\mathbb R^{\mathcal S\times \mathcal R}$ is the
	$(s,r)$ \textbf{entry} $\mathbb{M}^r_s\in\mathbb R$ of the matrix $\mathbb{M}$.

\item $\mathbb{M}^r$ is the $\mathcal S$ vector satisfying $(\mathbb{M}^r)_s=\mathbb{M}^r_s$ for all $r\in\mathcal R$.

\item $\mathbb{M}_s$ is the $\mathcal R$ vector satisfying $(\mathbb{M}_s)_r=\mathbb{M}^r_s$ for all	$s\in\mathcal S$.
\end{itemize}

\subsection{Chemical reaction networks}
\label{sec:CRN_intro}

A \textbf{chemical reaction network (CRN)} $\mathcal{G}=\{\Spc,\C,\R\}$ is a hypergraph\footnote{A graph is defined as a set of vertices and a set of edges linking the vertices. A hypergraph is a generalization of a graph with a set of hyperedges which can link multisets of vertices, also called hypervertices. (A multiset is a generalization of a set that allows for multiple instances of each element.)} 
consisting of three sets (see \cite{feinberg2019foundations}):
\begin{enumerate}
\item a set $\mathcal{S}$, elements of which are the \textbf{species} of the network (\textit{vertices} of the hypergraph).
\item a set $\mathcal{C}$ of $\mathcal{S}$ (column) vectors in $\mathbb{Z}_{\geq 0}^{\mathcal{S}}$, elements of which of multisets of species and called the \textbf{complexes} of the network (\textit{hypervertices} of the hypergraph).
\item a set $\mathcal{R} \subset \mathcal{C} \times \mathcal{C}$ such that (\textit{hyperedges} of the hypergraph)
\begin{enumerate}
\item for each $y \in \mathcal{C}$, $(y,y) \not\in \mathcal{R}$.
\item for each $y \in \mathcal{C}$ there is a $y' \in \mathcal{C}$ such that $(y,y') \in \mathcal{R}$ or $(y',y) \in \mathcal{R}$.
\end{enumerate}
\end{enumerate}

 Members of $\mathcal{R}$ are the \textbf{reactions} of the network. A reaction $r \in \mathcal{R}$ is of the form $(r^-,r^+)$, where $r^-$ and $r^+$ are the \textbf{input} and \textbf{output complexes}, respectively, in $\mathcal{C}$. In what follows, we use the notation $r^- \to r^+$ to denote $(r^-,r^+)$.

The \textbf{input} (\textbf{output}) \textbf{matrix} $\St^-$ ($\Sp$) is an $\Spc \times \R$ matrix whose columns are the input (output) complexes, $(\St^-)^r = r^-$ ($(\Sp)^r = r^+)$. The difference of the output and input matrices is called the \textbf{stoichiometric matrix} and denoted as $\St$, where $\St = \Sp-\St^-$. The columns of $\St$ are called the \textbf{reaction vectors} and the $r^\text{th}$ column is denoted as $\Delta r$, where $\Delta r = \St^r = (r^+-r^-).$

Define the $\mathcal{S} \times \mathcal{C}$ \textbf{complex stoichiometry matrix} $Y$, such that $Y^y = y \in \mathcal{C}.$ Since the CRN is a hypergraph, it is a graph over the complexes (hypervertices). Define the $\mathcal{C}\times \mathcal{R}$ \textbf{vertex-edge matrix} $\mathbb{M}$ such that
\[ \mathbb{M}_y^r = \begin{cases}
	+1 \text{ if } y = r^+,\\
	-1 \text{ if } y = r^-\\
	\phantom{+}0 \text{ otherwise.}
	\end{cases} \]
Then the stoichiometric matrix is given as the product of the matrix of complex stoichiometries and the vertex-edge matrix
$\mathbb{S} = Y \mathbb{M} .$

A \textbf{concentration vector} is an $\mathcal{S}$ vector $q \in \mathbb{R}^\mathcal{S}_{\geq 0}$ that consists of the concentrations of the species. 
An $\mathcal{R}$ (column) vector in the positive-orthant of the domain of the stoichiometric matrix $j \in \mathbb{R}_{\geq 0}^\mathcal{R}$ will be called a \textbf{reaction-flow vector}. 
The image of any reaction-flow vector under the stoichiometric matrix will be called a \textbf{species-flow vector}. Under a reaction-flow vector $j$, the rate of change of the concentration $q$ will be given by the species-flow vector $\dot{q}$ such that 
\begin{align}
\dot{q} &= \mathbb{S}j. \label{eq:rate_equation}
\end{align}
Note that $\dot{q}$ and $j$ have the same units.

Any $\mathcal{S}$ (column) vector $c^* \in \text{Ker}(\mathbb{S}^T)$ is called a \textbf{conservation law} \cite{feinberg2019foundations}. We denote the number of independent conservation laws or dimension of $\text{Ker}(\mathbb{S}^T)$ by $\ell$.

The image of the stoichiometric matrix $\text{Im}(\St)$ is called the \textbf{stoichiometric subspace}. By the rank-nullity theorerm, we know that the dimension of the stoichiometric subspace $\text{dim}(\text{Im}(\St)) = |\Spc|-\ell$. Two concentration vectors $q$ and $q'$ are \textbf{stoichiometrically compatible} if $q - q' \in \text{Im}(\St)$. Stoichiometric compatibility is an equivalence relation that induces a partition of $\mathbb{R}^{\Spc}_{\geq 0}$ into equivalence classes called the \textbf{stoichiometric compatibility classes} (Def.\ 3.4.6, \cite{feinberg2019foundations}).

Any reaction-flow vector $j \in \text{Ker}(\mathbb{S})$ is a \textbf{null flow}. Null flows can arise because either $j \in \text{Ker}(\mathbb{M})$ or $\mathbb{M}j \in \text{Ker}(Y)$. Any null flow in $\text{Ker}(\mathbb{M})$ and $\text{Ker}(Y)$ will be called a \textbf{null cycle} and $\mbf{\delta-}$\textbf{flow} \cite{smith2017flows}, respectively. We denote the dimension $\text{Ker}(\mathbb{M})$ and $\text{Ker}(Y)$ as $\iota$ and $\delta$, respectively. Elementary considerations yield that for a CRN $\{\mathcal{S},\mathcal{C},\mathcal{R}\}$ (App.\ A, \cite{gagrani2024polyhedral}), 
\begin{equation}
\delta = |\mathcal{R}| - |\mathcal{S}| - \iota + \ell. \label{eq:deficiency}    
\end{equation}
$\delta$ is called the \textbf{deficiency} of the network. Intuitively, the deficiency counts the number of independent null flows in the reaction network that are not visible from the graph of complexes but explicitly arise due to their stoichiometries.

\textbf{Mass-action kinetics (MAK)} is a \textit{kinetic model} (Def.\ 1, \cite{vassena2024unstable}) for a CRN $\{\mathcal{S},\mathcal{C},\mathcal{R}\}$ where:
\begin{enumerate}
\item each reaction $r \in \mathcal{R}$ is assigned a \textbf{rate constant} $k_r \in \mathbb{R}_{>0}$. 
\item for a given concentration vector $q$, a reaction-flow vector $j$ is assigned such that for each reaction $r: r^- \to r^+ \in \R$, 
$j_r = k_r q^{r^-}$.
\end{enumerate} 
Thus, under MAK, the dynamics of the concentration vectors $q$ is given by the ODES
\begin{equation}
    \dot{q}(q) = \sum_{r \in \R}\mathbb{S}^r j_r = \sum_{r \in \R}\mathbb{S}^r k_r q^{r^-}. \label{eq:MAK_eqs}
\end{equation} 
We denote the set of all rate constants as $\mathcal{K}$, and henceforth specify a CRN under MAK with the quadruple $\{\mathcal{S},\mathcal{C},\mathcal{R},\mathcal{K}\}$. A concentration $\eq$ is an \textbf{equilibrium} concentration of the system if $$\dot{q}(\eq) = 0.$$

\subparagraph{Reversible CRNs:}
A CRN is called \textbf{reversible} if for every forward reaction $r:r^- \to r^+ \in \mathcal{R}$, there exists its reverse reaction $r': r^+ \to r^- \in \mathcal{R}$.
For reversible CRNs, we employ the following notation. First, we obtain a unique set of reactions $\R_u \subset \R$ which contains either the forward or reverse reaction for each reaction in $\R$, but not both. We denote the $\Spc \times \R_u$ stoichiometric matrix $\St$ and reaction $r \in \R_u$. For each $r$, we denote the forward and reverse rate constants as $k_r^+$ and $k_r^-$, respectively, forward and reverse reaction-flow vectors as 
\[j_r^+ = k_r^+ q^{r^-} \text{ and  } j_r^- = k_r^- q^{r^+},\]
respectively, the net reaction-flow vector through $r$ as 
\[j_r = j_r^+ - j_r^-,\]
and the resulting concentration-flow vector as
\begin{equation}
    \dot{q} = \sum_{r\in \R_u} (\St)^r j_r = \sum_{r\in \R_u} (\St)^r (j_r^+-j_r^-). \label{eq:reversible_MAK}
\end{equation}

In this work, for reversible CRNs, we use the following strong version of the theorem due to Horn, Jackson, and Feinberg (Theorem 7.1.1, \cite{feinberg2019foundations}).  
\begin{theorem}[The Deficiency Zero Theorem]
\label{thm:def_zero}
    A reversible CRN of deficiency zero taken under MAK, regardless of the rate constants, has precisely one stable equilibrium within each stoichiometric compatibility class.
\end{theorem} 
\noindent Furthermore, it can be shown that the equilibrium for deficiency zero reversible CRNs is either \textbf{detailed-balanced} (Ch.\ 14, \cite{feinberg2019foundations}), where the forward and backward reaction-flow for each reaction are identical
    \begin{align}
        j_r(\eq) &= 0, \nonumber\\
        j_r^+(\eq) &= j_r^-(\eq), \nonumber\\
       k_r^+ \eq^{r^-} &= k_r^- \eq^{r^+},  \label{eq:detailed_balanced}
    \end{align}
or \textit{complex-balanced} (see Ch.\ 15, \cite{feinberg2019foundations} for details).
 
\begin{example}
\label{eg:emb_gen_schlogl_defnt}
We define a class of CRNs, parametrized by $g \in \mathbb{Z}_{> 0} $, that we refer to as the \textbf{complete generalized-Schlogl model} and denote as $\mathcal{M}'_g$. Taken under MAK, $\mathcal{M}'_g = \{\mathcal{S}'_g,\mathcal{C}'_g,\mathcal{R}'_g,\mathcal{K}'_g\}$ where
\begin{align*}
\mathcal{S}'_g &= \{ X, E_0, E_2, E_3, \ldots, E_{2g+1}\},\\
\mathcal{C}'_g &= \{ X, E_0, E_2 + E_0 + 2X, 3X + E_3,  \ldots,\\
& \phantom{=0} E_{2g} + E_0 + (2g)X, (2g+1)X + E_{2g+1}\},\\
\{\mathcal{R}'_g,\mathcal{K}'_g\} &= \{\ce{E_0 <=>[$k_0'$][$k_1'$] X},\\
& \phantom{=0} \ce{E_2 + E_0 + 2X <=>[$k_2'$][$k_3'$] 3X + E_3},\\
 & \phantom{=0} \ldots \\
 & \phantom{=0}  \ce{E_{2g} + E_0 + (2g)X <=>[$k_{2g}'$][$k_{2g+1}'$] (2g + 1)X + E_{2g+1}}\}.
\end{align*}

Let us denote the stoichiometric matrix for $\mathcal{M}'_g$ as $\mathbb{S}'_g$. Then for $g=2$,
\[  \mathbb{S}'_2 = 
\begin{bmatrix}
\phantom{-}1 & -1 & \phantom{-}1 & -1 & \phantom{-}1 & -1 \\
-1 & \phantom{-}1 & -1 & \phantom{-}1 & -1 & \phantom{-}1\\
\phantom{-}0 & \phantom{-}0 & -1 & \phantom{-}1 &\phantom{-}0 & \phantom{-}0\\
\phantom{-}0 & \phantom{-}0 & \phantom{-}1 & -1 &\phantom{-}0 & \phantom{-}0\\
\phantom{-}0 & \phantom{-}0 &\phantom{-}0 & \phantom{-}0 & -1 & \phantom{-}1 \\
\phantom{-}0 & \phantom{-}0 &\phantom{-}0 & \phantom{-}0 & \phantom{-}1 & -1 \\
\end{bmatrix}
 \]
For any $g$, we have
\begin{align*}
|\mathcal{S}'| & = 2g + 2,\\
|\mathcal{R}'| & = 2g +2,\\
\iota &= g+1,\\
\ell & = g+1.
\end{align*}
$\iota$ can be counted by observing that any null cycle is given by going forward and backward on the same reaction vector (and these are the only null cycles). $\ell$ can be counted by observing that the pairs of species with the same conserved quantity are $\{X, E_0\}$, $\{E_{2i},E_{2i+1}\}_{i \in [1,g]}$ (and these are the only conservation laws).
This yields that for any $g$, the deficiency of $\mathcal{M}_g'$ is zero, $\delta (\mathcal{M}_g') = 0$. Thus, applying Theorem \ref{thm:def_zero}, every CRN in the complete generalized-Schl\"{o}gl model class has exactly one asymptotically stable equilibrium irrespective of the choice of rate constants. 
\end{example}

\begin{example}
\label{eg:gen_schlogl_defnt}
We define another class of CRNs, parametrized by $g \in \mathbb{Z}_{> 0} $, that we refer to as the \textbf{generalized-Schlogl model} \cite{lazarescu2019large} and denote as $\mathcal{M}_g$. Taken under MAK, $\mathcal{M}_g = \{\mathcal{S}_g,\mathcal{C}_g,\mathcal{R}_g,\mathcal{K}_g\}$ where
\begin{align*}
\mathcal{S}_g &= \{ X\},\\
\mathcal{C}_g &= \{\vn, X, 2X, \ldots,(2g+1)X\},\\
\{\mathcal{R}_g,\mathcal{K}_g\} &= \{\ce{$\vn$ <=>[$k_0$][$k_1$] X},\\
& \phantom{=0} \ce{2X <=>[$k_2$][$k_3$] 3X },\\
 & \phantom{=0} \ldots \\
 & \phantom{=0}  \ce{(2g)X <=>[$k_{2g}$][$k_{2g+1}$] (2g + 1)X}\}.
\end{align*}

Let us denote the stoichiometric matrix for $\mathcal{M}_g$ as $\mathbb{S}_g$. Then for $g=2$,
\[  \mathbb{S}_2 = 
\begin{bmatrix}
\phantom{-}1 & -1 & \phantom{-}1 & -1 & \phantom{-}1 & -1 
\end{bmatrix}
 \]
For any $g$, we have
\begin{align*}
|\mathcal{S}| & = 1,\\
|\mathcal{R}'| & = 2g +2,\\
\iota &= g+1,\\
\ell & = 0.
\end{align*}
$\iota$ can be counted by observing that any null cycle is given by going forward and backward on the same reaction vector (and these are the only null cycles). $\ell = 0$ as there are no conservation laws.
This yields that for any $g$, the deficiency of $\mathcal{M}_g$ is $g$, $\delta (\mathcal{M}_g') = g$. Thus, we cannot apply Theorem \ref{thm:def_zero} and this CRN may exhibit multiple equilibria. 
\end{example}

\subsection{Stochastic chemical kinetics}
\label{sec:stoch_chemical_kin}
The discrete nature of CRNs is modeled stochastically as a pure jump Markov process where a state is given by a species \textbf{population vector} ($n \in \mathbb{Z}^{\Spc}_{\geq 0}$) \cite{anderson2015stochastic}. A reaction, when it occurs, replaces an input combination of species sampled without-replacement from the system into an output combination and the system jumps to a new state. An object of study for stochastic CRNs is the probability distribution $\mathbb{P}(n,t)$ of finding the system in state $n$ at time $t$. It is well-known that this probability distribution evolves under the chemical master equation \cite{ge2013chemical} and the dynamics of its first moment (average) reduces to the MAK in systems with a very large number of particles \cite{kurtz1972relationship}.

Denoting the volume of the system by $V$, consider the stochastic dynamics of the process $q^V_t = n(t)/V$ describing the concentration of species at time $t$ starting from some concentration $q^V_0$. 
From path-integral methods (Ch.\ 4, \cite{smith2015symmetry}; Ch.\ 4, \cite{kamenev2023field}), it is known that the probability of observing $q^V$ at time $t$ conditioned on the system being in $q^V_0$ at time $0$ is asymptotically given by the large-deviation (LD) principle of the form (Eq.\ 226, \cite{touchette2005legendre})
\begin{equation}
     \mathbb{P}(q_t^V|q^V_0) \asymp e^{-V S(q_t,q_0)}, \label{eq:LDF_escape}
\end{equation}
where the rate function $S(q_t,q_0)$ is evaluated along the optimal trajectory with appropriate boundary conditions
\[ S(q_t,q_0) = \inf_{q(t):q(0)=q_0,q(t)=q_t} J[q]\]
of the functional 
\begin{equation}
    J[q] = \sup_{p(t)} \int_0^t \,dt \left( p \cdot \dv{q}{t} - H(p,q) \right). \label{eq:J_rate}
\end{equation}
Here, $H(p,q)$ is the \textbf{Hamiltonian} of the process and for CRNs, it takes the form
\begin{align}
    H(p,q) &= \sum_{r \in \mathcal{R}} (e^{(r^+-r^-)\cdot p}-1)k_{r}q^{r^-}, \nonumber\\
        &= \sum_{r \in \mathcal{R}} (e^{\Delta r \cdot p}-1)k_{r}q^{r^-},\label{eq:Ham-kinetics}
\end{align}
where $p \in \mathbb{R}^\mathcal{S}$ is the conjugate momentum. For more on variational methods in LD theory and LD theory for CRNs, see App.\ C in \cite{shwartz1995large} and \cite{agazzi2018large}, respectively.

The rate function can be re-expressed as value along the saddle-point solutions
\[S(q_t,q_0) = \inf_{q(t):q(0)=q_0,q(t)=q_t} \sup_{p(t)} \mathcal{A}[p,q] \]
of the \textbf{action functional} 
\begin{equation}
    \mathcal{A}[p,q] = \int \,dt \left( p \dv{q}{t} - H(p,q) \right). \label{eq:action_func}
\end{equation}
It is then straightforward to show that the rate function satisfies the \textbf{Hamilton-Jacobi PDEs}  \cite{goldstein2002classical}
\[ \pdv{S}{t} = -H\left(\pdv{S}{q_t},q_t\right).\]
The \textbf{stationary distribution} of a stochastic process, if it exists, is the probability distribution over states (or its corresponding rate function) that does not change with time. It can be obtained from stochastic simulations \cite{gillespie1977exact,gillespie2007stochastic} by simulating the system for a very long time and normalizing the \textbf{residence times} of the system in (a small neighbourhood of) each state. Since the probability of a state $q_t$ in the stationary distribution is independent of the initial condition and time, the stationary distribution is only a function of the state where evaluated and we denote it as $\pi(q)$. Denote the rate function of the stationary distribution, also called the \textbf{non-equilibrium potential (NEP)} \cite{anderson2015lyapunov} by $\V(q)$. Thus, the NEP must satisfy   
\begin{equation}
    H\left(\pdv{\V}{q},q\right) = 0. \label{eq:NEP}
\end{equation}
Thus, the NEP is calculated by finding solutions to the Hamilton-Jacobi equations in the $H=0$ submanifold (Sec.\ 4.10, \cite{kamenev2023field}).

It is well-known that the integral curves to the Hamilton-Jacobi PDEs are the solutions of \textbf{Hamilton's equations of motion (EoM)}
\begin{align}
\dv{q}{t} &= \phantom{-}\pdv{H(p,q)}{p}, \nonumber\\
\dv{p}{t} &= -\pdv{H(p,q)}{q}.
    \label{eq:HamEoM}
\end{align}
Since, in this work, we are only interested in understanding the stationary distribution, we only investigate the classes of solutions of Hamilton's EoM in the $H=0$ submanifold, namely: relaxation and escape trajectories \cite{smith2020intrinsic}.

The equations of MAK (compare with Eq.\ \ref{eq:MAK_eqs}) are obtained as the $p=0$ solution 
\begin{align}
    \dot{q}_\text{rel} := \dv{q}{t}\bigg|_{p=0} &= \pdv{H}{p}\bigg|_{p=0}\nonumber\\
    &= \sum_r (r^+-r^-)k_{r}q^{r^-}, \nonumber\\
    &= \sum_r \St^r k_{r}q^{r^-} \label{eq:relax_Ham}
\end{align}
and their solutions are called \textbf{relaxation trajectories} of the system. 

The \textbf{fixed points} (or equilibrium) of the relaxation trajectories, when they exist, inform about the long-term behavior dynamics of CRNs. Denoting an arbitrary fixed point as $\eq$ and their set as $\{\eq\}$, we have $\dot{q}_\text{rel}(\eq)=0.$
The stability of a fixed-point can be determined by evaluating the eigenvalues of the \textbf{Jacobian matrix} $J$, where
\[J = \pdv{\dot{q}_\text{rel}}{q} = \pdv{H}{q}{p}\bigg|_{p=0},\]
at that point. In particular, we term fixed-points where all non-zero eigenvalues have negative real parts as \textbf{attractors}, all but one non-zero eigenvalues have negative real parts as \textbf{transient points}, and all other fixed-points as \textbf{unstable points}. CRNs can also exhibit periodic or chaotic attractors, but in this work we only consider CRNs with (point) attractors as defined above. A system with multiple attractors is called \textbf{multistable}. 

The solutions with $p \neq 0$ in the $H=0$ submanifold are termed as the \textbf{escape trajectories}. These trajectories are in the phase space (concentrations and their conjugate momenta) that  and we denote them as $(q_\text{esc},p_\text{esc},t)$. The projection of the escape trajectories on the concentration space give the least-improbable (or most-probable) paths out of the attractor to a point. In a stochastic simulation, these one expects these to be the most probable paths for the system to make an excursion out of the attractor in the large volume limit, and the value of the action along them gives an estimate of the escape probability (Eq.\ \ref{eq:LDF_escape}).

While any two points $q_1$ and $q_2$ may not be joined together by a single escape trajectory, it is possible to join them with a concatenation of escape trajectories with the time variable possibly running against the direction of the curve parametrization (Sec.\ 3.2.2, \cite{smith2020intrinsic}), and we denote it as $(q^*,p^*,t^*)$ where $q^*(0) = q_1$ and $q^*(\max(t^*)) = q_2$. Using Eqs.\ \ref{eq:J_rate} and \ref{eq:NEP}, the difference in NEP between the two points is given as 
\begin{align}
    \mathcal{V}(q_2) - \mathcal{V}(q_1) &= \int_{q^*} \pdv{\V}{q} \cdot \,dq = \int_{q^*} p^*(q) \cdot \,dq. \label{eq:NEP_diff}
\end{align}
For an information-geometric explanation of the role of escape trajectories in estimating the NEP, see Sec.\ IV A 5 \cite{smith2023rules}.

A transient point is \textbf{adjacent} to an attractor if there is a relaxation trajectory that starts in the vicinity of the transient point and terminates at the attractor. By the mountain pass theorem \cite{evans2022partial,bisgard2015mountain}, it is known that any two \textit{nearby} attractors must have at least one transient point in between them. Two attractors adjacent to the same transient point will be called adjacent to each other. A multistable stochastic CRN spends indefinite amount of time in the vicinity of an attractor before transitioning out of it due to fluctuations and entering another attractor adjacent to it. For reversible CRNs, wherein for every reaction, its reverse is also possible, there must exist trajectories, particularly an optimal or \textit{escape} trajectory, that emanate out of the attractor and terminate at the transient point. Estimating these escape trajectories can be computationally challenging and can be carried out by using the shooting-method \cite{dykman1994large} or functional gradient descent algorithms as proposed in \cite{gagrani2023action} or \cite{zakine2023minimum}. Once determined, they can then be used in Eq.\ \ref{eq:NEP_diff} to estimate the difference in the NEP at the attractors (Sec.\ 4.2.2, \cite{smith2020intrinsic}).

Suppose a transient point $t$ has two adjacent attractors $a_1$ and $a_2$. The following theorem (Lemma 1, \cite{smith2020intrinsic}) proves that, for a multistable CRN, the NEP at attractors is lesser or equal than that at the adjacent transient point
\[ \V(t) \geq \V(a_1), \hspace{0.2em} \V(t) \geq \V(a_2).\]
\begin{theorem}
    The NEP is a Lyapunov function along the relaxation trajectories given by MAK. 
    \label{thm:Lyapunov}
\end{theorem}
\begin{proof}
    First, recall that the equations of MAK are the gradient of the Hamiltonian in the momentum variable evaluated at $p=0$,
    \[\dot{q}_\text{rel} = \pdv{H}{p}\bigg|_{p=0}.\]
    Moreover, the Hamiltonian is identically zero at $p=0$
    \[H(0,q) = 0.\]
    Secondly, notice that the Hessian of the Hamiltonian in the momentum variables is positive-definite,
    $$\forall v \in \R^{\Spc}, \quad v^T\pdv{H}{p^T}{p}v \geq 0,$$
    or equivalently the Hamiltonian is convex in the momentum variables. This means that for any concentration $q$, 
    \[H(p,q) \geq H(0,q) + p \cdot \pdv{H}{p}\bigg|_{p=0}.\]
    Since the escape momentum $p_\text{esc}$ lies in the $H=0$ submanifold, we have
    \[H(p_\text{esc},q) = 0 \geq 0 + p_\text{esc} \cdot \pdv{H}{p}\bigg|_{p=0}, \]
    which yields
    \[p_\text{esc} \cdot \pdv{H}{p}\bigg|_{p=0} \leq 0 \hspace{0.25em} \forall q.\]
    Finally, using Eqs.\ \ref{eq:relax_Ham} and \ref{eq:NEP_diff}, we get
    \[\pdv{\mathcal{V}}{q} \cdot \dot{q}_\text{rel} \leq 0,\]
    which concludes the proof.
\end{proof}

  \begin{figure*}
        \centering
        \begin{subfigure}[b]{0.475\textwidth}
            \centering
            \includegraphics[width=\textwidth]{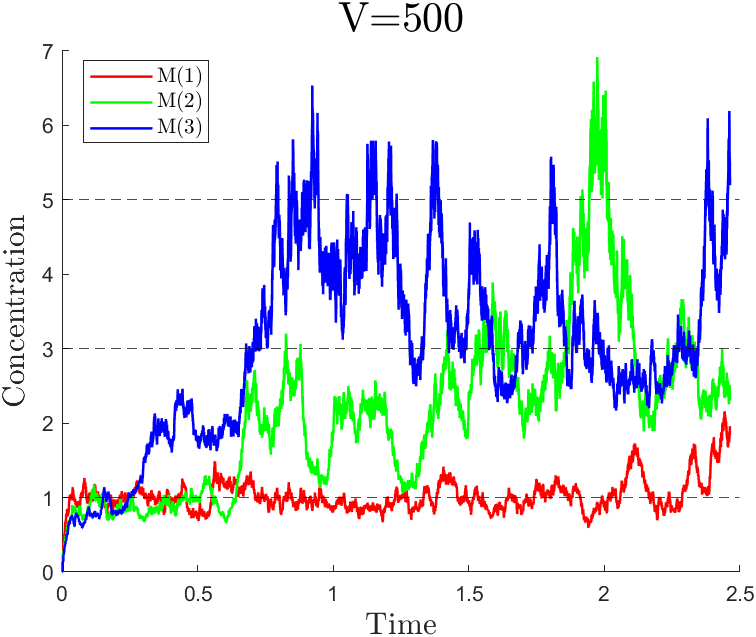}
            \caption[Network2]%
            {{\small Stochastic simulation}}    
            \label{fig:mean and std of net14}
        \end{subfigure}
        \hfill
        \begin{subfigure}[b]{0.475\textwidth}  
            \centering 
            \includegraphics[width=\textwidth]{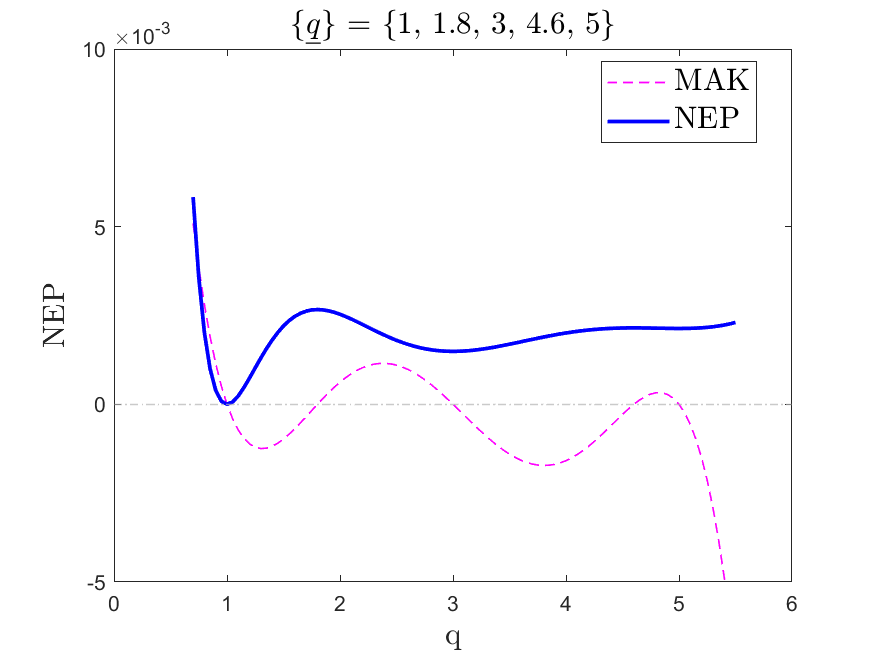}
            \caption[]%
            {{\small Model 1}}    
            \label{fig:mean and std of net24}
        \end{subfigure}
        \vskip\baselineskip
        \begin{subfigure}[b]{0.475\textwidth}   
            \centering 
            \includegraphics[width=\textwidth]{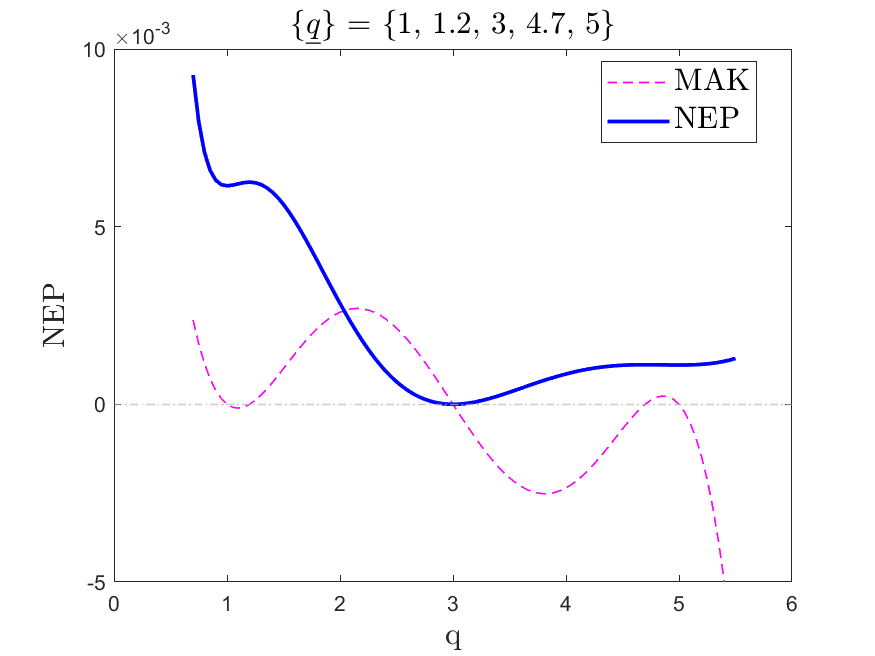}
            \caption[]%
            {{\small Model 2}}    
            \label{fig:mean and std of net34}
        \end{subfigure}
        \hfill
        \begin{subfigure}[b]{0.475\textwidth}   
            \centering 
            \includegraphics[width=\textwidth]{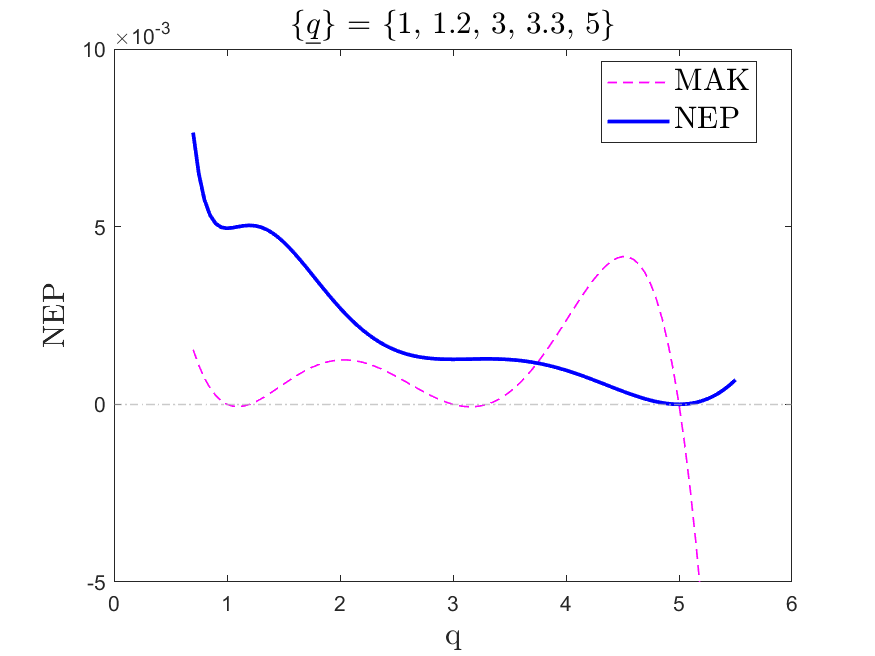}
            \caption[]%
            {{\small Model 3}}    
            \label{fig:mean and std of net44}
        \end{subfigure}
        \caption[]
        {Stochastic simulation using Gillespie algorithm, non-equilibrium potential (NEP), and mass-action kinetic (MAK) polynomial is showed for the three models from Example \ref{eg:ex4_3models_gen_schlogl}. The MAK polynomial is arbitrarily scaled to overlay on the NEP to indicate the positions of the fixed-points and their stability.} 
        \label{fig:gen_schlogl_sim_NEP}
    \end{figure*}

\begin{example}

For a CRN with a detailed balanced (Eq.\ \ref{eq:detailed_balanced}) or complex-balanced equilibrium, it can be shown that (see App.\ B, \cite{gagrani2023action}):
\begin{align}
    \pdv{\V}{q} = p_\text{esc} &= \ln\left(\frac{q}{\eq}\right), \nonumber\\
    \V(q) - \V(\eq)  &= \int_{\eq}^q  \ln\left(\frac{q}{\eq}\right) \cdot \,dq \nonumber\\
    &= q\ln\left(\frac{q}{\eq}\right) - (q-\eq). \label{eq:NEP_comp_balanced}
\end{align}
This a well-known result for complex-balanced zero deficiency CRNs and the NEP $\V$ is also called the \textit{Horn-Jackson potential} \cite{horn1972general,gopalkrishnan2013lyapunov}. Since the complete generalized Schl\"{o}gl model from E.g.\ \ref{eg:emb_gen_schlogl_defnt} is a deficiency zero reversible CRN and has a unique detailed-balanced or complex-balanced equilibrium for any choice of rate constants, its NEP must be of the form \ref{eq:NEP_comp_balanced}. 
\end{example}

\begin{example}
\label{eg:ex4_3models_gen_schlogl}
Consider the generalized Schl\"{o}gl model from E.g.\ \ref{eg:gen_schlogl_defnt} with one species and non-zero deficiency. Observe that there are only two distinct reaction vectors in the CRN $\Delta r = \pm 1$. Every reaction with $\Delta r=+1$ or $-1$ can be seen as a birth or death process, respectively, and the CRN is a \textit{birth-death} process \cite{anderson2015lyapunov}. 

Define polynomials $f_\pm(q)$ (suppressing the dependency on rate constants) with the rates of birth and death as follows: 
\begin{align*}
    f_+(q) &= \sum_{i=0}^{g} k_{2i} q^{2i},\\
    f_-(q) &= \sum_{i=0}^{g} k_{2i+1} q^{2i+1}.
\end{align*}
Using Eq.\ \ref{eq:Ham-kinetics}, the Hamiltonian for $\mathcal{M}_g$ is
\[H(p,q) = (e^p-1)f_+(q) + (e^{-p}-1)f_-(q). \]
Using Eq.\ \ref{eq:relax_Ham}, the relaxation trajectory or MAK equation is
\[\dot{q}_\text{rel} = f_+(q)-f_-(q) =\sum_{i = 0}^{2g +1} (-1)^i k_i x^i. \]

The escape solution for the system is obtained as the $p \neq 0$ solution along the $H=0$ submanifold. Since the phase space is two-dimensional, one constraint uniquely identifies $p(q) \neq 0$ in the $H=0$ submanifold, yielding 
\[p_\text{esc} = \log{\left(\frac{f_-(q)}{f_+(q)}\right)}.\]
One can verify that $H(p_\text{esc},q)=0$ for all $q$ and the this solution satisfies Eq.\ \ref{eq:HamEoM}.
Since this is a one-dimensional system, the escape trajectories in concentration space are trivial, and the difference of the NEP between any two points is given by 
\begin{align*}
    \mathcal{V}(q_2) - \mathcal{V}(q_1) &= \int_{q_1}^{q_2} \log{\left(\frac{f_-(q)}{f_+(q)}\right)} \,dq.
\end{align*}

For this model, as an application of Vi\'{e}te's formula \cite{vinberg2003course}, there is a bijection between rate constants and fixed points.
Suppose the rate constants were such that $\dot{q}_\text{rel}$ had $2g+1$ positive real fixed points,
\[ 0\leq \eq_1 \leq \ldots \leq \eq_{2g+1}.\]
Then, we can also express the relaxation trajectory as 
\[\dot{q}_\text{rel} = - \prod_{i=0}^{2g+1} (q - \eq_i).\]

For concreteness, we consider three models $\mathcal{M}^{(i)}_2, i \in \{1,2,3\}$ with $g=2$. The set of rate constants $\mathcal{K} = \{k_0,k_1,k_2,k_3,k_4,k_5\}$ are given as:
\begin{align*}
    \mathcal{K}^{(1)} &= \{124.2, 286.44, 236.72, 88.88, 15.4, 1\},\\
    \mathcal{K}^{(2)} &= \{84.6, 218.22, 201.46, 81.74, 14.9, 1\},\\
    \mathcal{K}^{(3)} &= \{59.4, 158.58, 154.14, 67.46, 13.5, 1\}.    
\end{align*}
The fixed-points for these models are:
\begin{align*}
    \{\eq\}^{(1)} &= \{ 1,1.8,3,4.6,5\},\\
    \{\eq\}^{(2)} &= \{ 1,1.2,3,4.7,5\},\\
    \{\eq\}^{(3)} &= \{ 1,1.8,3,3.3,5\}.
\end{align*}
A stochastic run for each of these models and their NEPs overlaid with (arbitrarily scaled) MAK polynomial can be found in Fig.\ \ref{fig:gen_schlogl_sim_NEP}. Notice that the relative positions of the roots determine the \textit{phase} of the dynamics and affect of the adjacent attractors has a higher residence time.    
\end{example}

\subsection{Thermodynamics of CRNs}
\label{sec:thermo_CRN}

For reversible CRNs, the \textbf{entropy production rate (EPR)} $\sigma$ is defined as (Eq.\ 55, \cite{rao2016nonequilibrium}) 
\begin{equation}
    \sigma(j) =  \sum_{r\in \R_u} \ln\left(\frac{j^+_r}{j^-_r}\right)j_r, \label{eq:EPR}
\end{equation}
where $\R_u, j^+_r, j^-_r$ are defined above Thm.\ \ref{thm:def_zero} in App.\ \ref{sec:CRN_intro}. In this subsection, we will restrict to reversible CRNs and henceforth use $\R$ to mean $\R_u$. Observe that each term of the type $\log(a/b)(a-b)$ is non-negative, and thus $\sigma(j) \leq 0$ for all $j$. In fact, $\sigma(j)=0$ only at detailed-balanced equilibrium where $j^+_r = j^-_r$. 

A CRN will be said to be \textbf{thermodynamically consistent} if each species can be assigned a \textbf{chemical potential of formation} $\mu_f \in \mathbb{R}^\mathcal{S}$ such that each reaction $r \in \R_u$ can be assigned a barrier height $E^b_r$ and the rate constants are of the form
\begin{align}
    k_r^+ &= e^{-(E^b_r - \mu_f \cdot r^-)/(k_BT)}, \nonumber\\
    k_r^- &= e^{-(E^b_r - \mu_f \cdot r^+)/(k_BT)}, \nonumber
\end{align}
where $k_B$ and $T$ are the Boltzmann constant and temperature, respectively. In what follows, we choose units where $k_B T = 1$. We can re-express the condition in terms of the ratio of $K = k_r^+/k_r^-$, also called the \textbf{equilibrium constant}, as 
\begin{equation}
   \ln K_r := \ln\left( \frac{k^+_r}{k^-_r}\right) = -\mu_f \cdot (r^+-r^-). \label{eq:equilibrium_constant}
\end{equation}
It is known from a generalization of Wegscheider's conditions \cite{schuster1989generalization} that $\ln K \in \text{Im}(\St^T)$, as in the above equation, if and only if there is a detailed balanced equilibrium $\eq$ that satisfies Eq.\ \ref{eq:detailed_balanced}
\[  K_r = \frac{\eq^{r^+}}{\eq^{r^-}}. \]
On comparing the above two equations, we find that 
\[ \St^T (\mu_f + \ln \eq) = 0.\]

Define the \textbf{equilibrium chemical potential}  $$\mu_0 := \mu_f + \ln \eq,$$ and the \textbf{chemical potential} 
\[ \mu :=  \mu_f + \ln(q) = \mu_0 + \ln\left(\frac{q}{\eq}\right). \]
This yields the \textbf{affinity} 
\begin{align}
  \ln\left(\frac{j^+_r}{j^-_r}\right) &= -(\mu_f+\ln(q)) \cdot (r^+-r^-) \nonumber\\ 
  &= - \mu \cdot (r^+-r^-), \label{eq:affinity}
\end{align}
and the EPR (Eq.\ \ref{eq:EPR}) becomes
\begin{equation}
    \sigma = -\mu^T \St j = -\ln\left(\frac{q}{\eq}\right) \cdot \dot{q}, \label{eq:balanced_EPR}
\end{equation}
where we use $\mu_0 \St = 0$ and $\dot{q}$ to refer to the MAK flow from Eq.\ \ref{eq:reversible_MAK}. The chemical potential determines the gain in free energy to the system by adding a particle of the species at a concentration (Sec.\ III C, \cite{rao2016nonequilibrium}). Thus, the affinity measures the loss in free energy due to one occurrence of a reaction at that concentration, and the EPR is rate of loss in free energy for the complete CRN. Define the \textbf{entropy produced (EP)} as the integral of the EPR in time 
\begin{equation}
    \Sigma = \int \,dt \sigma. \label{eq:EP}
\end{equation} 
Observe that the EP is negative of the the Horn-Jackson potential or the NEP for detailed balanced CRNs (Eq.\ \ref{eq:NEP_comp_balanced}). Thus, from the Lyapunov property of NEP, the EP by a thermodynamically consistent CRN never decreases as the system evolves deterministically.  

It is known that CRNs with a detailed-balanced or complex-balanced equilibrium have precisely one equilibrium in each stoichiometric compatibility class (Theorems 14.2.3 and 15.2.2, \cite{feinberg2019foundations}). While Feinberg's characterization of \textit{quasi-thermodynamic kinetic systems} (Sec.\ 13.4, \cite{feinberg2019foundations}) admits both detailed-balanced and complex-balanced systems, only CRNs with a detailed-balanced equilibrium admit a thermodynamic description as described above. Eq.\ \ref{eq:affinity} is used as a starting point  in Sec.\ III, \cite{rao2016nonequilibrium} to define the \textbf{Gibb's free energy of reaction} 
\begin{equation}
  \Delta_r G := \mu^T \mathbb{S}^r =  -\ln\left(\frac{j^+_r}{j^-_r}\right), \label{eq:Gibbs_reaction}  
\end{equation}
also called the \textit{thermodynamic force} driving reaction $r$, to define the thermodynamics of CRNs. For reaction flow $j_r \in \mathbb{R}$, the EPR per reaction $r$ is 
\begin{equation}
    \sigma_r(j_r) = - \Delta_r G j_r, \label{eq:EPR_reaction}
\end{equation}
and the EPR of the CRN is given by their sum over all reactions
\begin{equation}
    \sigma = \sum_r \sigma_r. \label{eq:EPR_CRN_2}
\end{equation}
For various thermodynamic decompositions of EPR for complex-balanced and multistable CRNs, see \cite{yoshimura2023housekeeping}.

The considerations above demonstrate that in a (thermodynamically consistent) \textbf{closed system}, where the dynamics of all species is modeled by MAK, the concentration will always approach a detailed-balanced equilibrium. However, in \textbf{open systems}, where the dynamics of only a species subset is modeled by MAK whereas the others are controlled by the environment, the system can be driven towards and maintained at a \textbf{non-equilibrium steady state (NESS)}. 

Henceforth, we follow terminology and notation from Sec.\ \ref{sec:internal-external}. 
Assuming that there are no reactions in the complete CRN not in $\R$, i.e., $\notR = \vn$, the \textbf{complete stoichiometric matrix}
\[ \nabla = \begin{bmatrix}
    \nabla^{\R}_\mcl{E} \\
    \St
\end{bmatrix}.\]
If the open system is allowed to evolve for a sufficiently long time, it will relax to a NESS assuming that the internal CRN has point attractors. The species-flow through the boundary species at the NESS is not necessarily zero and is the source of thermodynamic work done by the environment to maintain the NESS. Any non-vanishing boundary species-flow at the NESS is called an \textbf{emergent cycle} \cite{polettini2014irreversible}, denoted by $\epsilon \in \mathbb{R}^{\R}$, and satisfies
\begin{equation*}
    \nabla \epsilon = \begin{bmatrix}
    \nabla^{\R}_\mcl{E} \epsilon \\
    0
\end{bmatrix}.
\end{equation*}
The basis of emergent cycles (Sec.\ V, \cite{hirono2021structural})
$$\{\epsilon\} = \text{Ker}(\St)/\text{Ker}(\nabla)$$ 
consists of all reaction-flows in the kernel of the internal stoichiometric matrix that are not in the kernel of the embedding stoichiometric matrix. Observe that the EPR due to $\epsilon$ only depends on the chemical potential of the external-boundary species (Sec.\ II D, \cite{smith2023rules})
\begin{equation}
\sigma(\epsilon) = -\mu^T \nabla \epsilon = - \sum_{s \in \mcl{E}} \mu_s (\nabla^{\R}_\mcl{E} \epsilon)_s. \label{eq:EPR_emergent}    
\end{equation}

From physical considerations, it is useful to decompose the complete CRN $\mcl{G}'$ through a basis of \textbf{processes} \cite{wachtel2022free} $$p\in \mathcal{P}: p^- \to p^+ ,$$
that respect the conservation laws and span the stoichiometric subspace. 
Thus, analogously to the stoichiometric matrix, define the \textbf{process matrix} $\mathbb{P}$ such that
\begin{align*}
    (\mathbb{P})^p &= p^+-p^-,\\
    \text{Ker}(\mathbb{P}^T) &= \text{Ker}(\nabla^T),\\
    \text{Im}(\mathbb{P}) &= \text{Im}(\nabla).
\end{align*}
To transform from the reaction basis to process basis, we define a $\mcl{P} \times \R$ \textbf{process-reaction matrix} $\mbf{C}$ such that 
\[ \sum_p (\mathbb{P})^p_s \mbf{C}_p^r = \nabla^r_s \hspace{1em} \forall s \in \Spc'. \]

Using unprimed variables for reactions and primed variables for processes, analogous to Eq.\ \ref{eq:Gibbs_reaction}, we define \textbf{Gibb's free energy of process} 
\[\Delta'_p G := \mu^T (\mathbb{P})^p. \]
 For reaction flow $j \in \mathbb{R}^{\R}$, the flux through process $p$ $j'p \in \mathbb{R}^\mathcal{P}$ is
 \[j'_p = \sum_r \mbf{C}_p^r j_r.\]
 The EPR for a process for a given reaction-flow $j$
\begin{align}
    \sigma'_p(j) &=  - \mu^T (\mathbb{P})^p (\mbf{C})_p j, \nonumber \\
    &= -\Delta'_p G  j'_p, \label{eq:process_EPR}
\end{align}
and the total EPR (analogous to Eq.\ \ref{eq:EPR_CRN_2}) is
\begin{equation}
\sigma = \sum_{p \in \mathcal{P}} \sigma'_p. \label{eq:EPR_CRN_3}    
\end{equation}

From the definition of EPR in Eq.\ \ref{eq:EPR}, we know each term is individually nonnegative or a reaction always dissipates chemical potential energy. However, under a process decomposition of a reaction, the EPR of a process, $\sigma'_p$ in Eq.\ \ref{eq:process_EPR}, can be negative in which case the process will be said to be doing chemical work or \textbf{transducing free energy} at the expense of other processes which dissipate chemical potential at a higher rate to ensure that the total EPR in Eq.\ \ref{eq:EPR_CRN_3} is positive. The transduction \textbf{efficiency} of the processes is defined as (Sec.\ V, \cite{wachtel2022free})  
\begin{equation}
    \eta = - \frac{\text{Output}}{\text{Input}}, \label{eq:efficiency}
\end{equation}
where 
\begin{align*}
    \text{Output} &= \left\{\sum_p \sigma'_p \bigg| \sigma'_p < 0 \right\}, \\
    \text{Input} &= \left\{\sum_p \sigma'_p \bigg| \sigma'_p > 0 \right\}. \\
\end{align*}
Using Eqs.\ \ref{eq:process_EPR} and \ref{eq:efficiency}, we can calculate the efficiency of the environment in maintaining a NESS at emergent cycle $\epsilon$ with EPR given by Eq.\ \ref{eq:EPR_emergent}.

  \begin{figure*}
        \centering
        \begin{subfigure}[b]{0.49\textwidth}  
            \centering 
    \includegraphics[width = \textwidth]{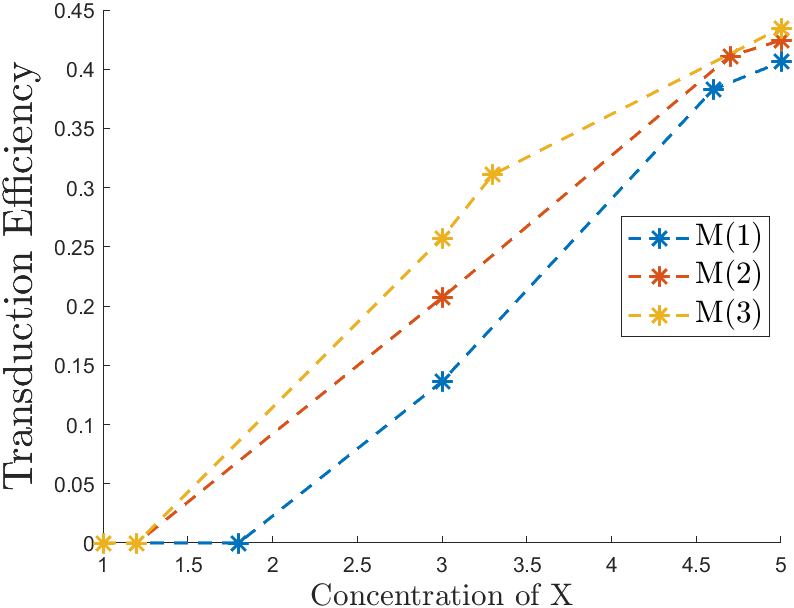}
        \end{subfigure}
        \hfill
           \begin{subfigure}[b]{0.49\textwidth}
            \centering
    \includegraphics[width = \textwidth]{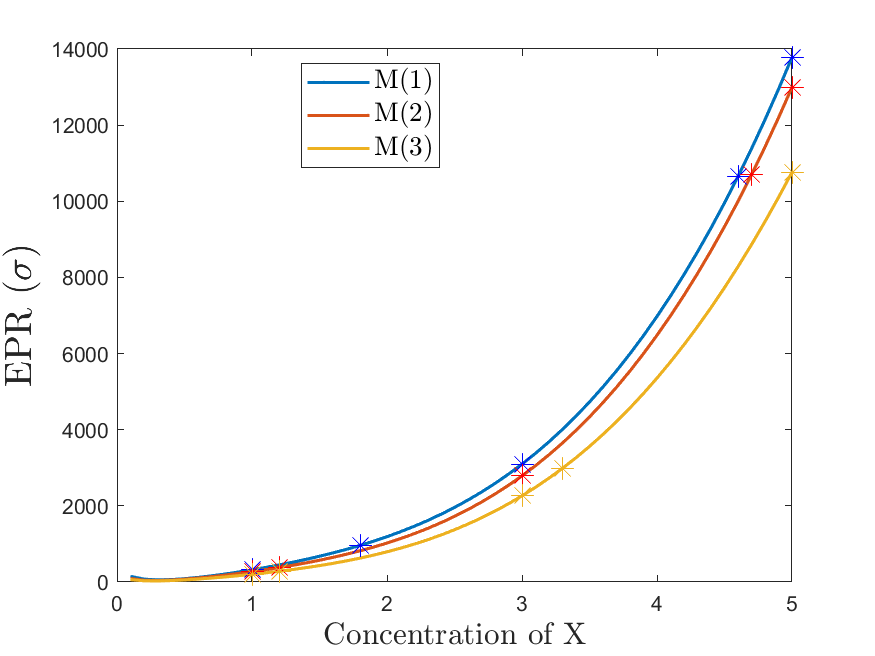}
        \end{subfigure}
        \caption[]
        {The transduction efficiency at the fixed points and EPR for all concentrations of the 3 models from E.g.\ \ref{eg:ex4_3models_gen_schlogl} are shown in the left and right panel, respectively. For these models, the attractors are fixed at \{1,2,3\} and the transient points in between vary. While the NEP profile for these models are different (Fig.\ \ref{fig:gen_schlogl_sim_NEP}), the transduction efficiency and EPR monotonically increase in the distance of the fixed points from the origin.}
    \label{fig:transduction_efficiency}
    \end{figure*}

\begin{example}
\label{eg:schlogl_efficiency}
 Consider the complete generalized-Schlogl model from E.g.\ \ref{eg:emb_gen_schlogl_defnt} as the embedding CRN. Since it is a reversible zero deficiency CRN and every linkage class has a single reaction, the equilibrium must be detailed-balanced and hence the embedding CRN is thermodynamically consistent. Let the boundary set $$\Spc_B = \{E_0, E_2, \ldots , E_{2g + 1}\}$$ chemostatted at concentrations $q^*$ by the environment.
 The resulting internal CRN then is the generalized Schlogl model from E.g.\ \ref{eg:gen_schlogl_defnt} with rate constants
 \begin{align*}
      k_0 &= k_0' q^*_{E_0}, \\
      k_1 &= k_1',\\
      k_{2j} &= k_{2j}' q^*_{E_0} q^*_{E_{2j}}, \\
      k_{2j+1} &= k_{2j+1}' q^*_{E_{2j+1}},
 \end{align*}
for $j \in [1,g].$ 

Recall that the unique reactions in the embedding CRN are
\begin{align*}
    \mathcal{R} &= \{r_0: \ce{E_0 -> X},\\
& \phantom{=0} r_1: \ce{E_2 + E_0 + 2X -> 3X + E_3},\\
 & \phantom{=0} \ldots \\
 & \phantom{=0} r_g: \ce{E_{2g} + E_0 + (2g)X -> (2g + 1)X + E_{2g+1}}\}.
 \end{align*}
For this system, we define the set of processes $\mcl{P}$ to consist of
\[p_0 := \ce{E_0 -> X}, \hspace{1em} p_j := \ce{E_{2j} -> E_{2j+1}},\]
for $j \in [1,g].$ 
Observe that the set respects the conservation laws and spans the stoichiometric subspace of the embedding CRN. It is easy to see that 
\begin{equation}
    r_0 = p_0,\hspace{1em} r_j = p_j + p_0 \label{eq:process_decomp_eg}
\end{equation}
for $j \in [1,g].$ This can be used to obtain the process-reaction matrix $\mbf{C}$ as defined earlier in the subsection.

To write each reaction's change in Gibb's energy and EPR, and decompose it using processes, observe that we can write the MAK as a sum over reaction pairs as 
\begin{align}
    \dot{q} &= \sum_{j=0}^g  k_{2j+1} \left(\ell_j - q \right)q^{2j}, \label{eq:MAK_ellj}
\end{align}
where 
\[  \ell_j := \frac{k_{2j}}{k_{2j+1}} \text{ for } j \in [0,g].\]
Using Eq.\ \ref{eq:Gibbs_reaction}, the change in Gibb's energy for the $j^\text{th}$ reaction is
\begin{align*}
    \Delta_j G &= \ln \left( \frac{q}{\ell_j}\right).
\end{align*}
The above equation, along with the decomposition of reactions into processes from Eq.\ \ref{eq:process_decomp_eg}, yields 
\begin{align*}
    \Delta_0' G &= \Delta_0 G = \ln \left( \frac{q}{\ell_0}\right),\\
    \Delta_j' G &= \Delta_j G - \Delta_0' G = \ln \left( \frac{\ell_0}{\ell_j}\right),
\end{align*}
for $j \in [1,g]$.
For a choice of rate constants where all fixed points are real, using \cite{branden2015unimodality}, the sequence of the coefficients $k_i$ is log-concave, or 
\[ k_i^2 \geq k_{i-1} k_{i+1}.\]
This yields,
\begin{align*}
    k_i^2 k_{i-1}^2 &\geq k_{i-1} k_{i+1} k_{i-2} k_{i}\\
    k_i k_{i-1} &\geq k_{i+1} k_{i-2} \\
    \ell_{i} & \geq \ell_{i-1},
\end{align*}
and, in particular, 
\begin{equation}
    \ell_0 \leq \ldots \leq \ell_g. 
\end{equation}
Thus, 
\[ \Delta_j' G  < 0, \text{ for } j \in [1,g].\]

The set of NESSs to which the open system can relax are given by the fixed-points of the internal CRN. 
Let the set of fixed points $\{\eq\}$ of the MAK of the internal CRN in species $X$ satisfy
\[ 0 \leq \eq_1 \leq \ldots \leq \eq_{2g+1},\]
such that 
\[ \dot{q} = - \prod_{i=0}^{2g+1} (q - \eq_i).\]
At any of the fixed points, the current through process $0$ that involves $X$ must be zero ($j'_0=0$), and only the remaining processes produce entropy. Using Eq.\ \ref{eq:process_EPR}, the EPR through process $j \in [1,g]$ at a fixed point $\eq$ is 
\[ 
\sigma_j'(\eq) =  -\left(\Delta'_j G\right)  k_{2j+1} \left(\ell_j - \eq \right)\eq^{2j}.
\]

As shown above, $\Delta'_j G$ is always negative for all $j$. For a process, different regimes of entropy production and free-energy transduction will be achieved when $\eq$ crosses $\ell_j$. In particular, process $j$ produces entropy when $\eq < \ell_j$ and transduces free energy when $\eq > \ell_j$. This means that more processes transduce free energy rather than produce entropy as we move from a fixed point closer from the origin to one that is further. Thus, we \textit{conjecture} that the model's transduction efficiency at NESSs is monotonic in their distance from the origin.

The transduction efficiency at the fixed-points of the 3 models from E.g.\ \ref{eg:ex4_3models_gen_schlogl} are shown in the left panel of Fig.\ \ref{fig:transduction_efficiency}. Recall that the attractors for this model are at $\{1,3,5\}.$ Observe that the transduction efficiency is higher at a fixed point further away from the origin, and we conclude that for these models, the attractor further from the origin transduces free energy more efficiently. Also, from Eq.\ \ref{eq:EPR}, the total EPR at any concentration $q$ is (see right panel in Fig.\ \ref{fig:transduction_efficiency})
\[\sigma(q) = \sum_{j=0}^g -\ln\left(\frac{q}{\ell_j}\right)  k_{2j+1} \left(\ell_j - q \right)q^{2j}.\]
Observe that the entropy produced also increases monotonically at the fixed points in their distance from the origin, and we conjecture that the EPR at fixed points is monotonic in the distance of the fixed points from the origin for any choice of rate constants for this model. The implications of this conjecture are discussed in Sec.\ \ref{sec:remarks_EPR_evolution}.

\end{example}
%\end{comment}

\bibliographystyle{unsrt} 
\bibliography{bibliography}

\end{document}